\documentclass{ametsocV6.1}
\usepackage{natbib}
\usepackage{verbatim}
\usepackage{soul}
\newcommand{\de}{\mathrm{d}}
\newcommand{\DD}{\mathrm{D}}
\newcommand{\pe}{\partial}

\newcommand{\ve}[1]{\boldsymbol{#1}}
\newcommand{\qdt}[1]{\quad \mbox{#1} \quad}
\newcommand{\mean}[1]{\langle {#1} \rangle}
\DeclareMathOperator{\erf}{erf}
\DeclareMathOperator{\csch}{csch}

\DeclareMathOperator{\sgn}{sgn}
\newcommand{\pl}{\textsuperscript{+1}}
\newcommand{\Ro}{\{\varepsilon\}}
\newcommand{\Ub}{\left\{\text{Bu}^{-1}\right\}}
\newcommand{\Bu}{\left\{\text{Bu}\right\}}
\title{Next-order balanced model captures submesoscale physics and statistics}

\authors{Ryan Sh\`iji\'e D\`u,\aff{a}\correspondingauthor{Ryan Sh\`iji\'e D\`u, ryan\_sjdu@nyu.edu}
K. Shafer Smith,\aff{a} 
Oliver B\"uhler,\aff{a}}

\affiliation{\aff{a}{Center for Atmosphere Ocean Science, Courant Institute of Mathematical Sciences, New York University}}

\abstract{Using nonlinear simulations in two settings, we demonstrate that QG\pl, a potential-vorticity based next-order-in-Rossby balanced model, captures several aspects of ocean submesoscale physics. In forced-dissipative 3D simulations under baroclinically unstable Eady-type background states, the statistical equilibrium turbulence exhibits long cyclonic tails and a plethora of rapidly-intensifying ageostrophic fronts. Despite that the model requires setting an explicit, small value for the fixed scaling Rossby number, the emergent flows are nevertheless characterized by vorticity and convergence {values larger than the local Coriolis frequency}, as observed in upper-ocean submesoscale flows. Simulations of QG\pl~under the classic strain-induced frontogenesis set-up show realistic frontal asymmetry and a {provable} finite time blow-up, quantitatively comparable to simulations of the semigeostrophic equations.  The inversions in the QG\pl~model are straightforward {linear Poisson} problems, allowing for the reconstruction of all flow fields from the PV and surface buoyancy, while avoiding the semigeostrophic coordinate transformation. {Taken together}, these results suggest QG\pl~{might be} a useful tool for studying upper-ocean submesoscale dynamics.}

\begin{document}
\maketitle

% \epigraph{Besides, if the mystery of the symmetriads is ever solved, there’ll still be the asymmetriads...}{Stanisław Lem, \textit{Solaris}}

\section{Introduction}

The oceanic mesoscale is characterized by flows with small Rossby number ($\varepsilon$) and lateral scales near the deformation radius, placing them squarely in the parameter regime described by the quasigeostrophic (QG) model. Thanks to its simple, elegant and practical structure, the QG model has thus served as the primary theoretical framework through which mesoscale flows have been understood, including through observational analysis \citep[e.g.][]{gill_energy_1974,Stammer97,CheltonEtAl2007Global}, development of theories to predict mesoscale eddy heat and tracer fluxes \citep[e.g.][]{HeldLarichev_96,GalletFerrari_21}, their parameterization in models \citep[e.g.][]{TreguierEtAl97,MarshallEtAl12}, and many more examples.  In the last 20 years, extensive observational campaigns and high-resolution numerical simulations have revealed that oceanic scales about a decade below the mesoscales --- the submesoscales --- are populated by $O(1)$-Rossby-number ageostrophic features that fall outside of QG dynamics {\citep{ShcherbinaEtAl_13,Mcwilliams16}}.  These include, for example, an asymmetric distribution of vortices with a slightly anticyclonic mean and a long tail of intense cyclostrophic cyclones, and many rapidly-forming strongly convergent fronts.  The fronts, in particular, play a role in Earth's climate, as their associated with strong vertical velocities flux heat, carbon and oxygen into the ocean's interior \citep[e.g.][]{Ferrari11,BalwadaEtAl_21}, and nutrients from the deep into the mixed layer \citep[e.g.][]{TaylorOceanFrontsTrigger2011,LevyEtAl18}.

An enormous advantage of QG is that it is based on the advection of a single {dynamically active} tracer field --- a linear approximation to the Ertel potential vorticity (PV) --- from which all other dynamic variables can be found through an elliptic inversion.  The QG model is thus straightforward to simulate and analyze, and it naturally omits inertial and gravity waves. These filtered ``unbalanced''  waves move at much higher frequencies than the ``balanced'' motions described by QG, further strengthening the rationalization for filtering them from the equations.  Submesoscale fronts and cyclones, by contrast, move and evolve on time scales of days to hours, making the boundary separating them from unbalanced waves murky and debated.  In lieu of a clean way to separate their dynamics, simulations of the full primitive equations has remained the main tool of their investigation. 

In spite of the challenge in precisely delineating balanced from unbalanced motion at submesoscales, a model that omits fast waves, yet includes both QG dynamics and ageostrophic non-wave submesoscale structures, would {likely} be useful for the interpretation of submesoscale observations \citep[e.g. from {Surface Water and Ocean Topography (SWOT)} altimetry,][]{morrowGlobalObservationsFineScale2019}, for developing submesoscale parameterizations in numerical climate models, and more.  Where should one start in developing or choosing a candidate model of this type?

There is a long history to the development of balanced models beyond QG \citep[see, e.g.][for general discussions]{McWilliamsGent_80,Vanneste13}.   Some of {these theoretical efforts were} aimed at the practical problem of creating computationally efficient gravity-wave-omitting models for simulation of the atmosphere and ocean at scales ranging from the deformation scale to the global scale.  Others were concerned with the understanding the so-called ``slow manifold'' \citep{Lorenz80,Leith80} --- an idealization in which slow, PV-conserving dynamics exist entirely independently of gravity waves --- as well as loss of balance \citep[e.g.][]{Vanneste13}.  Semigeostrophic (SG) dynamics, while originally devised as a two-dimensional model specifically to describe the ageostrophic secondary circulation that gives rise to rapid frontogenesis, was later generalized to a three-dimensional balanced model, distinguished by using a transformation to a frame of motion moving with the geostrophic flow \citep{HoskinsBretherton_72,Hoskins_75,Hoskins_82a}.  Semigeostrophy has served as a useful interpretive model for submesoscale frontogenesis \citep{ThomasEtAl_08a}, but simulating its nonlinear evolution is extremely challenging, requiring {at} each time step the solution of a nonlinear Monge–Ampère equation, as well as coordinate frame transformations \citep[see][for a laudable example of its simulation in {a surface-dominated-setting}]{RagoneBadin_16}. Moreover, it is in fact not higher-order in Rossby number than QG \citep{McWilliamsGent_80}.

The Balance Equations \citep[BE,][]{Charney_55a,McWilliamsGent_80} have also been used to study some aspects of submesoscale dynamics. For example, \citet{Grooms_15a} used the BE to study the ageostrophic effects on the linear stability of the Eady mean state. The BE also conserves energy and can be modified to conserve PV \citep{Allen_91}. The BE has two time derivatives, but are nevertheless balanced in the sense that there is only one true prognostic variable. The resolution of this apparent paradox lies in the fact that the two time evolution equations are in fact not independent, but are connected in complex and nonlinear ways and cannot be reduced to one equation. This makes evolving the model in time not straightforward and sometimes impossible \citep{McWilliamsGent_80}. 

Another branch in the development of balanced models proceeds by asymptotically expanding the primitive equations in a small parameter (e.g. the Rossby number) to second order.  \cite{WarnEtAl_95} argued on theoretical grounds that when doing so, one variable --- termed, unfortunately, the ``slaving" variable --- must be left unapproximated in order to avoid secular growth.  
\citet[hereafter V96]{Vallis_96a} develops a collection of such models, some expanded about small Rossby number, and using the potential vorticity (PV) as the unapproximated variable, ensuring that the PV is exactly conserved. However, these models do not guarantee energy conservation.
While a number of models are derived in V96, only cases with distinguished limits appropriate for large-scale flow, and only in a shallow water setting, are investigated numerically. 

The choice of using PV as the starring variable is well-motivated by its historically central role in facilitating understanding of geophysical fluid dynamical processes, including ocean frontal processes \citep{Hoskins_74,HaynesMcIntyre_87,MarshallNurser_92,Thomas_05}. The idea of obtaining all other physical variables from PV alone is also well-accepted (cf. `invertibility principle' in \citealp{HoskinsEtAl_85}).
However, such models do not guarantee energy conservation (in contrast to, for example, Hamiltonian-based models, e.g. \citealp{Salmon98}).
One remark is warranted for the energy conservation in the SG model. The energy that it conserves only includes the geostrophically balanced components of the velocities \citep{Hoskins_75}, thus making utilizing its energy budget {subtle} for submesoscale-focused studies.

\citet[hereafter MSR99]{MurakiEtAl_99} revisited the approach of V96, but proceeded by first introducing a reformulation of the primitive equations through the use of an elegant vector potential.   The resulting so-called QG\textsuperscript{+} equations are then expanded in Rossby number to the next order past QG, resulting in the so-called QG\pl~model.  While equivalent to one of the models derived in V96 (see Appendix~A), the specific formulation devised by MSR99 allows easy inclusion of boundary conditions (which V96 neglects) and is more computationally tractable.  Like QG, the QG\pl~model consists of the advection of a single PV variable but requires additional elliptic inversions to obtain ageostrophic components of the flow.  These ageostrophic corrections remove QG's shackle of symmetry, allowing the model to capture richer ocean submesoscale phenomena. 
{We note that in a companion work, we explore the shallow water version of QG\pl/V96 model in further detail \citep{DuEtAl_24_SWQGp1}.}

The linear dynamics of QG\pl~under a baroclinically unstable background was studied in an atmospheric context by \cite{RotunnoEtAl_00,MurakiHakim_01}.  To date, few investigations of its nonlinear dynamics have been published \citep{HakimEtAl_02,Weiss_22,MaaloulyEtAl_23}. \citet{HakimEtAl_02,MaaloulyEtAl_23} considered the model in a surface-dynamics setting, {named SQG\pl}. It is akin to the SQG model \citep{HeldEtAl95}, with surface buoyancy advection bounding a semi-infinite, constant-PV interior.  The freely-evolving simulations of {\citet[][hereafter HSM02]{HakimEtAl_02}} demonstrate that the model produces significant beyond-QG features, including asymmetries in vorticity, fronts, and filaments, as well as surface restratification, all reminiscent of atmospheric near-surface mesoscale dynamics, as well as in upper-ocean submesoscale flows.  We {also note} the recent work by \cite{MaaloulyEtAl_23}, who simulate SQG\pl~forced by white-in-time noise, demonstrating that the model captures realistic horizontal particle dispersion behavior. \citet{Weiss_22} studies the point vortex solutions of QG\pl~in triply periodic domain, {showing that} high-order corrections introduce asymmetric behavior.

Motivated by the relative ease with which it can be numerically simulated, and its ability to capture features observed in upper-ocean submesocale flows, here we simulate the fully nonlinear QG\pl~dynamics in two key settings. 
% using the pseudospectral solver Dedalus for all our simulations \citep{BurnsEtAl_20}. 
The first is the classic case of steady-state turbulence driven by an unstable background Eady mean flow, which also serves as a model for submesoscale Mixed Layer Instability \citep[MLI,][]{BoccalettiEtAl_07}. Simulations in this setting reveal a flow with kinematic statistics matching those found in submesoscale flows \citep{BalwadaEtAl_21} to a remarkable degree.  Notably, although QG\pl~retains an explicit Rossby number $\varepsilon$ multiplying the second-order corrections, the emergent flow exhibits much larger flow-Rossby numbers --- specifically, though the explicit Rossby number is set to $\varepsilon = 0.032$, the emergent flow contains some vortices with flow-Rossby number as high as $\zeta^\text{t}/f \sim 7$ (where $\zeta^\text{t}$ is the vertical component of vorticity at the sea surface, and $f$ is the Coriolis frequency). 

The second case considered is the strain-induced frontogenesis problem of \cite{HoskinsBretherton_72}. We show that the QG\pl~model captures the super-exponential growth of strain-induced fronts found in the SG model \citep{HoskinsBretherton_72,Hoskins_82a,ShakespeareTaylor_13}.  Moreover, we prove the existence of a finite-time blow-up in Appendix~B. 
Compared to the SG model, QG\pl~is asymptotically consistent and higher-accuracy in Rossby number, ensuring it represents curved fronts just as well as for straight fronts. Additionally, the inversion step in QG\pl~is numerically simpler, requiring only solving linear Poisson problems. Moreover, everything can be done in the Eulerian frame without a change of basis. Combined, we show that QG\pl~could be used to model the two most important processes that energize the submesoscale identified in modeling and parametrization studies \citep{ZhangEtAl_23a}. 

The paper is organized as follows.  In section~\ref{sec:QGpl_deriv} we review the main elements of MSR99's derivation of the QG\pl~model, connecting the essential steps to familiar concepts in the QG and primitive equations model. Section~\ref{sec:Eady} sets up the Eady-instability-forced formulation of QG\pl, and compares the statistics and features of its steady-state turbulence to that of the standard Eady-forced QG model.  Section~\ref{sec:2D_front} investigates the ability of QG\pl~to capture strain-induced frontogenesis in the classic cross-front vertical slice configuration. We compare its solutions to benchmark simulations of the SG model in the same configuration.  Finally, we conclude and speculate in Section \ref{sec:conclu}.

\section{Derivation of the QG\pl~model}
\label{sec:QGpl_deriv}

The QG\pl~model is derived in MSR99.  We repeat its main elements here to provide a self-contained presentation, and to clarify some issues that arise when trying to simulate the model in realistic configurations.  Specifically, we provide physical clarification for constant adjustments to the PV and surface buoyancies that are necessary for the elliptic inversions. 

\subsection{Scaled primitive equations and velocity potential}

The derivation begins with the hydrostatic Boussinesq equations (or primitive equations) with constant Coriolis frequency $f$ and Brunt--Väisälä frequency $N$ 
\begin{subequations}
\begin{align}
    \Ro\DD_t u -fv &= -p_x,\label{u}\\
    \Ro\DD_t v +fu &= -p_y,\label{v}\\
    p_z &= b,\label{w}\\
    \Ro\DD_t b +\Bu N^2w &= 0,\label{b}\\
    u_x+v_y+w_z &= 0.
\end{align}
\end{subequations}
Here $\ve{v} = (u, v, w)$ are the Cartesian velocity components, $p$ the pressure divided by the mean density, $b$ is the deviation from the total buoyancy, $b_\text{tot} = N^2z + \Ro b$, and $\DD_t = \pe_t + \ve{v}\cdot\nabla_3$ is the advective derivative with $\nabla_3$ the three-dimensional gradient.  The equations are presented simultaneously in dimensional and nondimensional forms, using the notation of V96: for the dimensional form, set the nondimensional constants in curly brackets to unity ($\varepsilon = \text{Bu} = 1$), and for the nondimensional form, set the dimensional constants unity ($f = N = 1$).  In nondimensional form, model variables are scaled following the standard QG scaling with the pressure scaling geostrophically (this is different from the scaling in MSR99 only if Burger number is not equal to one): $x,y \sim L$, $z \sim H$, $t \sim (\varepsilon f)^{-1}$, $u,v \sim U$, $w \sim U H/L$, $p \sim f U L$ and $b\sim f U L/H$, where $U$, $L$ and $H$ are typical scales of horizontal velocity, horizontal length, and vertical depth, respectively. The nondimensional parameters are the Rossby and Burger numbers
\begin{equation}\label{ndparams}
  \varepsilon := \frac{U}{fL} \quad\text{and}\quad
  \text{Bu}:= \left(\frac{NH}{fL}\right)^2.
\end{equation}

The primitive equations conserve the hydrostatic Ertel potential vorticity, which in dimensional form is $Q = \left[f\mathbf{\hat{z}}+\nabla_3\times(u,v,0)\right]
\cdot\nabla_3 b_\text{tot}$. 
%\begin{equation}\label{Q}
%  Q = \left(f\mathbf{\hat{z}}+\nabla_3\times\ve{v}_H\right)
%  \cdot\nabla_3 b_\text{tot}
%  = \nabla_3\cdot\left[\left(f\mathbf{\hat{z}}
%      +\nabla_3\times\ve{v}_H\right) b_\text{tot}\right] 
%\end{equation}
%where $\ve{v}_H = (u,v,0)$.
Expanding and including the nondimensional scalings gives
\begin{equation}
    Q = fN^2+\Ro q
\end{equation}
where
\begin{align}\label{PV} 
  q = \underbrace{N^2\zeta + \Ub fb_z}_{:= q_\text{QG}}
  + \left\{\frac{\varepsilon}{\text{Bu}}\right\}
  \nabla_3\cdot\left[\boldsymbol{\omega}b\right].
\end{align}
Here $\boldsymbol{\omega} =(-v_z,u_z,\zeta)$ and $\zeta = v_x-u_y$ are defined for convenience, and we note that
\begin{equation}\label{vortb}
    \nabla_3\cdot\left[\boldsymbol{\omega}b\right] = \zeta b_z-v_zb_x+u_zb_y.
\end{equation}
Because $N$ and $f$ are constant throughout this paper, the first term in $Q$ is constant, and the dynamically relevant PV, $q$, is the sum of the QGPV (when $N^2$ is constant) plus an ageostrophic quadratic correction. 

MSR99 first reconfigures the {primitive equations} in a form intended to facilitate the asymptotic approximation. Their approach begins with the introduction of three potential fields, $\Phi$, $F$, and $G$, as components of a vector potential to replace the three prognostic variables in \eqref{u}, \eqref{v} and \eqref{b} {(see (1) of \citet{MurakiEtAl_99} for a nondimensional expression)}.
% , namely $(v,-u,b) = \nabla_3 p +\Ro \left(\DD_t u,\DD_t v,0\right)$.
A more standard approach uses that any incompressible velocity field $\ve{v}$ can be written $\ve{v} = \nabla_3\times\ve{A}$ where $\ve{A}$  is a vector potential.  Setting $\ve{A}=(-G,F,-\Phi)$ (which differs by an overall minus from {(27) of} MSR99) gives
\begin{subequations}\label{velocity}
  \begin{align}
    u &= -\Phi_y-F_z,\label{upot}\\
    v &= \Phi_x-G_z, \label{vpot}\\
    w &= F_x+G_y. \label{wpot}
  \end{align}
\end{subequations}
{(the potentials} scale as $\Phi \sim UL$ and $G , F \sim UH$.)  In principle, if $\ve{A}$ is known, then the three prognostic and two diagnostic variables in the {primitive equations} should be uniquely determined through an inversion.  The connection of the velocities to the potential fields is explicit in \eqref{velocity}, but the relationship of the buoyancy $b$ and pressure $p$ to $\ve{A}$ remains to be shown.

The vector potential formulation  $\ve{v} = \nabla_3\times\ve{A}$  admits a gauge freedom in that the transformation
\begin{equation}\label{gauge}
    \ve{A} \to \ve{A} + \nabla_3\Gamma  
\end{equation}
leaves $\ve{v}$ unchanged. This freedom must be constrained if the correspondence between $\ve{A}$ and the {primitive equations} variables to be one-to-one.  One can do so by fixing the buoyancy $b$ to scaled version of $-\nabla_3\cdot\ve{A}$, resulting in 
\begin{align}\label{buoy}
    b &= f\Phi_z+\Bu \frac{N^2}{f} \left(G_x- F_y\right).
\end{align}
This choice has the advantage that the QGPV becomes
\begin{equation}\label{qgpv}
  q_\text{QG} = N^2\zeta +\Ub fb_{z} = \mathcal{L}(\Phi),
\end{equation}
where
\begin{equation}\label{Lop}
  \mathcal{L} = N^2\nabla^2 + \Ub f^2\pe_{zz}
\end{equation}
is a stretched version of the three-dimensional Laplacian and $\nabla = (\pe_x,\pe_y)$ is the horizontal gradient. 
Hence $\Phi$ is the only component of $\ve{A}$ that is relevant for the QGPV, which is convenient.  However, note that, unlike QG, the vorticity $f\zeta = \nabla^2\Phi + F_{yz} - G_{xz}$ depends on all three potential fields and $\Phi$ is not simply proportional to pressure.

It is also useful for future derivation to note that
\begin{align}
    \mathcal{L}(F) &= -\Ub f^2u_z-\Ub fb_{y}+N^2w_x,\label{eq:Lap_F}\\
    \mathcal{L}(G) &= -\Ub f^2v_z+\Ub fb_{x}+N^2w_y.\label{eq:Lap_G}
\end{align}
The stretched Laplacian of the three potentials also appears in \citet[][]{DeusebioEtAl_13}, where their equations (2.5) list their equations of evolution. We note these evolution laws only in passing to bolster their choice in QG\pl. But we will not attempt to evolve $F$ and $G$ and only look for ways to invert for them since we are looking for a balanced model with only one prognostic variable, the PV.

The gauge transformation \eqref{gauge} leaves $\ve{v}$ unchanged, but the buoyancy $b$ is unchanged only if
\begin{equation}\label{Gamma}
  \mathcal{L}(\Gamma)=0.
\end{equation}
Thus $\Gamma$ is harmonic in the stretched coordinates.  In a triply-periodic domain this implies that $\Gamma$ is constant and therefore the transformation from $(\ve{u},b)$ to $\ve{A}$ is unique.

With a rigid lid and flat bottom, the vertical velocity $w=0$ at the upper and lower boundaries of the domain, which from \eqref{wpot} requires
\begin{align}\label{barney}
    F_x + G_y = 0.
\end{align}
Non-constant gauges are possible now, and this can be used to simplify \eqref{barney} with the simpler and stronger
\begin{align}\label{barbie}
    F = G = 0.
\end{align}
The gauge transformation \eqref{gauge} implies $F \to F+\Gamma_y$ and $G \to  G-\Gamma_x$.  If $(F,G)$ have been found and satisfy \eqref{barney} at the boundaries, then solving \eqref{Gamma} with the upper and lower boundary conditions $F+\Gamma_y =0$ and $G-\Gamma_x =0$ means the transformed potentials satisfy \eqref{barbie}, as well as \eqref{barney}.  Such a $\Gamma$ is unique up to a linear function $m + n z$, which only adds an inconsequential constant $-n$ to $\Phi$. Hence we can use the stronger \eqref{barbie} and with this choice the transformation from $(\ve{v},b)$ to $\ve{A}$ is again unique.

With the gauge-transformed potential, \eqref{barbie} implies that
\begin{equation}\label{b_bc}
  b = f\Phi_z \quad\text{at the upper and lower boundaries}.
\end{equation}
However, as noted earlier, despite the resemblance of \eqref{b_bc} to the hydrostatic relation, the potential $\Phi$ is not simply proportional to the pressure.

It is worth emphasizing that so far no approximation has been made, so in principle one could rewrite the full {primitive equations} in terms of $\ve{A}=(-G,F,-\Phi)$ (cf. QG\textsuperscript{+} in MSR99 and \citet[(2.5)]{DeusebioEtAl_13}). For example, setting $\Phi=0$ means QGPV is zero and then suitably scaled $F$ and $G$ can describe linear inertia--gravity waves at leading order (Section 3c of MSR99 as well as Appendix B of \citet[][]{DeusebioEtAl_13}).

\subsection{Kinematic and dynamical constraints}

A key kinematic constraint arises from the fact that the Ertel PV can be written as a divergence.  From \eqref{PV} and \eqref{qgpv} the dynamical PV may be written as the three-dimensional divergence (in stretched coordinates)
\begin{equation}\label{PV2}
    q = \mathcal{L}(\Phi) 
    + \left\{\frac{\varepsilon}{\text{Bu}}\right\}\nabla_3\cdot 
    \left(\boldsymbol{\omega} b\right). 
\end{equation}
Integrating over the domain, assuming horizontally periodicity and using the improved surface boundary conditions \eqref{b_bc}, one has
\begin{equation}\label{PVintegral}
  \langle q\rangle = \left[f\overline{b}
  +\Ro\overline{\zeta b}\right]_{\text{b}}^{\text{t}}
\end{equation}
where 
\begin{equation}
  \left\langle \;\cdot\;\right\rangle
  := \frac{1}{{|V|}}\iiint_V \; \cdot \;\de x\de y\de z
  \quad\text{and}\quad
  \overline{(\;\cdot\;)}
  := \frac{1}{{|A|}}\iint_A \; \cdot \;\de x\de y
\end{equation}
are, respectively, the averages over the fluid volume $V$ and horizontal area $A$ evaluated at the top surface t or bottom surface b, and
\begin{align}
    \left[ \;\cdot\;\right]^\text{t}_\text{b} 
    = \left.(\;\cdot\;)\right|_{z=0}-\left.(\;\cdot\;)\right|_{z=-H}
\end{align}
the difference between the top and bottom. 
Equation \eqref{PVintegral} is a kinematic relationship that holds regardless of whether the flow is forced or dissipated.

We point out three dynamic integral constraints that hold in the absence of forcing and dissipation, two of which are specific to the surfaces. Because $w=0$ at either vertical boundary, the thermodynamic equation \eqref{b} becomes
\begin{equation}\label{bsurf}
   \pe_t b^\text{t,b} + \ve{u}^\text{t,b}\cdot\nabla b^\text{t,b} = 0,
\end{equation}
where $\ve{u}=(u,v)$ is the horizontal velocity vector. Taking the horizontal average gives
\begin{equation}\label{bbdivg}
    \pe_t \overline{b}^\text{t,b} = \overline{b^\text{t,b}\nabla\cdot\ve{u}^\text{t,b}}.
\end{equation}
Thus nonzero horizontal divergence --- which will be retained in QG\pl~--- allows for changes in mean surface buoyancy, which is driven by correlations of buoyancy and horizontal divergence.   

The second relation we derive implies the kinematic constraint \eqref{PVintegral}, but only in the absence of forcing and dissipation.
%, the above kinematic constraint is implied by a dynamical conservation. 
Taking the curl of \eqref{u}-\eqref{v}, and again using that $w=0$ at the upper and lower boundaries, gives the surface vertical vorticity equation
\begin{equation}
    \Ro \left(\pe_t\zeta^\text{t,b} + \ve{u}^\text{t,b}\cdot\nabla\zeta^\text{t,b}\right)  = \left(f+\Ro\zeta^\text{t,b}\right)w^\text{t,b}_z.
\end{equation}
%and the buoyancy equation is
%\begin{align}
%    \pe_t b^\text{t,b} + \ve{u}^\text{t,b}\cdot\nabla b^\text{t,b} = 0.
%\end{align}
Forming the sum $\Ro\overline{b\zeta_t} + \overline{\zeta b_t}$ and using integration by parts then gives a conservation law that applies separately at each surface,
\begin{equation}\label{surfconstraint}  
    \pe_t\left[f\overline{b^\text{t,b}} 
    +\Ro\overline{\zeta^\text{t,b} b^\text{t,b}}\right] = 0. 
\end{equation}
This relation is used in \citet{LapeyreEtAl_06} to study submesoscale frontogenesis.
Note that \eqref{surfconstraint} of course implies \eqref{PVintegral}, but is not necessarily stronger than the latter since \eqref{surfconstraint} only holds when there is no forcing or dissipation.

Furthermore, in the absence of forcing and dissipation, the {primitive equations} conserves energy, and this can be expressed using the potentials without approximation. The conserved energy is 
\begin{equation}\label{eq:total_energy}
    \text{E} = \text{EKE}+\text{APE},
\end{equation}
where
\begin{equation}\label{eq:EKE}
    \text{EKE} = \left\langle \frac{u^2+v^2}{2}  \right\rangle
\end{equation}
is the eddy kinetic energy and 
\begin{equation}\label{eq:APE}
    \text{APE} = \Ub\left\langle \frac{b^2}{2N^2} \right\rangle
\end{equation}
is the available potential energy.   Substituting in the potential form of the variables and using integration by parts with the boundary conditions \eqref{barbie}, one finds that all cross-terms cancel, and the total energy can be expressed 
\begin{align}\label{eq:TE_potform}
    \text{E} = \frac{1}{2}\left\langle |\nabla \Phi|^2
             + \Ub\frac{f^2}{N^2}\Phi_z^2+ F_z^2+G_z^2 +\Bu\frac{N^2}{f^2}(F_y-G_x)^2
             \right\rangle.  
\end{align}
The potential form decouples the total energy into a component involving only $\Phi$, which follows the form of the QG energy, and another that involves only $F$ and $G$ (cf. \citet{DeusebioEtAl_13}).

\subsection{The evolution and inversion equations of QG\pl}

QG\pl~is a balanced model whose only prognostic variables are the PV in the interior and buoyancy at the upper and lower surfaces. The inviscid equations are simply
\begin{subequations}\label{PVevolution}
    \begin{align}   
        \pe_t q + \ve{v}\cdot\nabla_3 q&= 0\label{eq:q_adv}\\
        \pe_t b^\text{t,b} + \ve{u}^\text{t,b}\cdot\nabla b^\text{t,b} &= 0.
    \end{align}
\end{subequations}
The rest of the derivation of the QG\pl~model involves finding ways to invert the physical variables from the prognostic variables by expanding asymptotically. 
Consistent with QG scaling, we assume 
\begin{align}
    \varepsilon \ll 1, \quad \text{Bu} = O(1),
\end{align}
and asymptotically expand $\ve{A}$ to first order in the small Rossby number as
\begin{align}\label{eq:var_expan}
\begin{split}
    u &= -\Phi^0_y-\Ro(\Phi^1_y+F^1_z),\\
    v &= \Phi^0_x+\Ro(\Phi^1_x-G^1_z),\\
    w &= 0+\Ro(F_x^1+G_y^1),\\
    b &= f\Phi^0_z +\Ro f\left(\Phi^1_z+\Bu \frac{N^2}{f^2} G^1_x
    - \Bu \frac{N^2}{f^2} F^1_y\right).
\end{split}
\end{align}
Here $F^0=G^0=0$ as these potentials represent ageostrophic motions.  

Using \eqref{eq:var_expan} to evaluate the PV \eqref{PV2}, one finds an expression for the dynamical PV up to $O(\varepsilon)$
\begin{align}\label{PV3}
    q = \mathcal{L}(\Phi^0) + \Ro \left[ \mathcal{L}(\Phi^1)+\Ub f \left(\Phi^0_{zz}\nabla^2\Phi^0-|\nabla\Phi^0_z|^2\right) \right] + {O}(\varepsilon^2).
\end{align}
Likewise, with \eqref{b_bc}, the surface buoyancies are 
\begin{align}\label{b3}
\begin{split}
  b^\text{t} &= \left.f\Phi_z^0\right|_{z=0}
  +\Ro \left.f\Phi_z^1\right|_{z=0}+ {O}(\varepsilon^2),\\
  b^\text{b} &= \left.f\Phi_z^0\right|_{z=-H}+\Ro \left.f\Phi_z^1\right|_{z=-H} + {O}(\varepsilon^2).
\end{split}
\end{align}
To invert the potentials, we impose the solvability condition of \citet{WarnEtAl_95} and V96 that, while the model advects the full dynamical Ertel PV \eqref{PV2} and surface buoyancies, each advected variable must, in fact, remain unexpanded (no Rossby order $q^1$, in contrary to the notation of MSR99 in style), in order to prevent secular growth.  We thus demand that at $O(\varepsilon^0)$, the potential $\Phi^0$ is given by inverting the elliptic equation
\begin{align}\label{eq:PV0_inv}
    q = \mathcal{L}(\Phi^0) + \mean{q},
\end{align}
with boundary conditions
\begin{equation}\label{eq:PV0_invbc}
  b^\text{t} = \left.f\Phi^{0}_z\right|_{z=0} + \overline{b}^\text{t},\qquad
  b^\text{b} = \left.f\Phi^{0}_z\right|_{z=-H} + \overline{b}^\text{b}.
\end{equation}
In the above, we've demanded that the constant offsets be given by the respective average of the prognostic field. This is motivated by the fact that the QG model does not allow for mean changes in PV and surface buoyancy, and ensures this elliptic problem with two Neumann boundary conditions is well-posed.

The principle of not expanding the PV, along with \eqref{PV3}-\eqref{eq:PV0_invbc} gives an elliptic equation for $\Phi^1$ in terms of $\Phi^0$,
\begin{equation}\label{eq:PV1_inv}
  \mathcal{L}(\Phi^1)
  = C_q-\Ub f \left(\Phi^0_{zz} \nabla^2\Phi^0-|\nabla\Phi^0_z|^2\right)
\end{equation}
with
\begin{equation}\label{eq:PV1_invbc}
  \Ro \left.f\Phi^{1}_z\right|_{z=0} = \overline{b}^\text{t},
  \quad\text{and}\quad
  \Ro \left.f\Phi^{1}_z\right|_{z=-H} = \overline{b}^\text{b}.
\end{equation}
This is again an elliptic problem with two Neumann boundary conditions, and $C_q$ is a free constant to ensure the problem is well-posed. Integrating \eqref{eq:PV1_inv} and using the boundary conditions \eqref{eq:PV1_invbc}, it needs to satisfy
\begin{align}\label{eq:q1constr1}
    \Ub f^2\left[\overline{\Phi^{1}_z}\right]^\text{t}_\text{b} &= C_q -\Ub f \left\langle \Phi^0_{zz}\nabla^2\Phi^0
    -|\nabla\Phi^0_z|^2\right\rangle\\
    &= C_q- \Ub f\left( \overline{\Phi^0_z\nabla^2\Phi^0} \right|^\text{t}_\text{b}.
\end{align}

This is the $O(\varepsilon)$ version of \eqref{PVintegral} if $C_q = \Ro\mean{q}$. For the inviscid case, $\mean{q}=0$ always because $q$ is advected by an incompressible velocity field, so the above is also implied by \eqref{surfconstraint} at $O(\varepsilon)$. However, the equivalence of $\Ro C_q$ and $\mean{q}$ is not guaranteed, as this implies a change in the average PV of the inverted physical fields. 
MSR99 attributes this to the asymptotic nature of the model. \eqref{PVintegral} and \eqref{surfconstraint} could have errors of $O(\varepsilon^2)$ when applied to the inverted field. A more important reason is that diffusion and dissipation are active in the real fluid, and necessary for numerical stability in any simulation. As a result, PV is not perfectly conserved, and source terms in the form of a flux (usually represented by a divergence of a $\ve J$ vector) must be introduced to the right-hand-side of \eqref{eq:q_adv}   \citep{HaynesMcIntyre_87,MarshallNurser_92}. 
In submesoscale flows, non-zero PV flux from the surface due to viscous effects can be important (cf. \citet{BodnerEtAl_19}). QG\pl~in its current form does not account for the viscous PV flux, and is therefore inconsistent with the need for diffusion in the surface buoyancy evolution. The constant $C_q$ is a crude representation of this PV flux,  accounting only for its effect on the mean. 
Even if the $\ve J$ term is small in the PV evolution equation, its effects will accumulate in the presence of sustained vigorous turbulence due to forcing (e.g. Eady baroclinic instability forcing in Section \ref{sec:Eady}). 

Though our choices of constant adjustments to satisfy the integral constraints are physically reasonable, a motivated reader might want to find another choice. For this, we remark that a constant change of the interior right-hand side and the boundary condition in the first order inversion \eqref{eq:PV1_inv} will not change the velocity fields because they only change a horizontally constant function to the solution $\Phi^1$. But changes in the zeroth order will. This means for our particular choices, the surface average buoyancy $\overline{b^\text{t,b}}$ does not effect the evolution of the model. 

To complete the model, we need inversion relations that relate the ageostrophic vertical streamfunctions $F^1$ and $G^1$ to $\Phi^0$. Our derivation is {inspired by that of the} $\omega$-equation by \cite{HoskinsEtAl_78a}. We start with the $O(\varepsilon)$ {primitive equations},
\begin{subequations}
\begin{align}
    \DD_t^0 u^0 - fv^1 &= -p^1_x,\label{eq:BouORo_u}\\
    \DD_t^0 v^0 + fu^1 &= -p^1_y,\label{eq:BouORo_v}\\
    \DD_t^0 b^0 + \Bu N^2w^1 &= 0,\label{eq:BouORo_b}\\
    p^1_z &= b^1.\label{eq:BouORo_hystat}
\end{align}
\end{subequations}
Taking the difference of the $z$-derivative of $f$ times \eqref{eq:BouORo_v} and the $x$-derivative of \eqref{eq:BouORo_b} gives
\begin{align}
  -\Ub& f^2u^1_z-\Ub fb^1_{y}+N^2w^1_x= \Ub \left[f \left(\DD_t^0 v^0\right)_z-\left(\DD_t^0b^0\right)_x\right],
\end{align}
where we also used \eqref{eq:BouORo_hystat}. The right-hand side simplifies due to thermal wind balance and becomes the familiar first element of the $\ve Q$ vector. The left-hand side is just the first order of \eqref{eq:Lap_F} and only depends on $F^1$. Together we have the inversion for $F^1$ to be a Poisson equation
\begin{align}
    \mathcal{L}(F^1) = \Ub 2f J(\Phi_z^0,\Phi_x^0). \label{eq:F_inv}
\end{align}
A similar derivation gives the inversion for $G^1$
\begin{align}
    \mathcal{L}(G^1) = \Ub 2f J(\Phi_z^0,\Phi_y^0).\label{eq:G_inv}
\end{align}
Both equations are solved with homogeneous Dirichlet boundary conditions at the surfaces \eqref{barbie}.

To summarize the PV inversion, given the prognostic variables $q$, $b^\text{t}$, and $b^\text{b}$, we can obtain the potentials $\Phi^0$, $\Phi^1$, $F^1$ and $G^1$ by solving four constant coefficient Poisson problems. The left-hand side elliptic operator is the same in each case, a typical situation for an asymptotic model. 

\subsection{A few physical implications of the model}

First, we note that QG\pl~ breaks the artificial symmetry imposed by QG.  Because of the quadratic terms in the right-hand sides of these Poisson problems, changing the sign of PV, we have
\begin{align}
\begin{split}
    q\to -q \qdt{$\Rightarrow$}& \Phi^0\to-\Phi^0,\\
    \qdt{but}& \Phi^1\to\Phi^1, F^1\to F^1, G^1\to G^1.
\end{split}
\end{align}
This is the feature of QG\pl~that introduces asymmetries beyond QG. 

Second, we show that QG\pl~captures a form of cyclogeostrophic balance.  To see this, it is useful to know first that the QG\pl~model is effectively equivalent to the balanced model for the ``rapidly rotating stratified primitive equations'' presented in V96 (Section 4(a)), which is based on ideas presented in \cite{WarnEtAl_95}. Because this is not obvious from MSR99, we show the connection explicitly in Appendix~A. Integrating \eqref{eq:V96_clyO1} in $z$ implies
\begin{align}
    \nabla^2 p^1-f\zeta^1 &= 2J(\Phi^0_x,\Phi^0_y).
\end{align}
Using the geostrophic relation $\nabla^2 p^0-f\zeta^0=0$ that holds at zeroth order, we find %for up to $O(\varepsilon)$
\begin{align}
    \nabla^2 p-f\zeta &= \Ro 2J(\Phi^0_x,\Phi^0_y)+O(\varepsilon^2).
\end{align}
This agrees with the cyclogeostrophic balance (or the divergence equation) up to $O(\varepsilon)$ (cf. (2.18) of \cite{McWilliamsGent_80}).

Lastly, we note explicitly an implication of the inclusion of frontogenesis in QG\pl.  In the {primitive equations} and in nature, frontogenesis and other ageostrophic processes act to restratify the surface ocean, changing the mean surface buoyancy, and concomitantly the subsurface mean interior buoyancy and hence the background stratification.  In the present setup, we define a fixed initial stratification $N^2$, and treat the leading term $fN^2$ in the definition of $Q$ as a constant.  However,  $\overline{b}^\text{t,b}$ can change (see \eqref{bbdivg}), and so a new stratification $N_\text{eff}^2 = N^2 + \overline{b}_z$ would evolve.   (In a perfectly inviscid simulation, the mean perturbation PV $\mean{q}$ would nevertheless remain constant, with changes in $\overline{b}$ compensated for by buoyancy-vorticity correlations as shown in \eqref{surfconstraint}). 

\section{QG\pl~forced by the Eady instability}
\label{sec:Eady}

The Eady instability \citep{Eady_49}, characterized by zero PV in the interior and two active buoyancy layers on the top and bottom, is an effective minimal model for baroclinic instability relevant to the ocean mixed layer, capturing the dynamical essence of the so-called mixed layer instability (MLI) \citep{BoccalettiEtAl_07}.
We use the Eady mean state to drive the QG\pl~system, studying the resulting turbulent steady state.
This is an extension of the SQG\pl~model studied in HSM02. Specifically, the Eady instability acts as a realistic board-band forcing to the dynamics, an aspect missed by the freely decaying simulations in HSM02, and improves on the white noise forcing in \cite{MaaloulyEtAl_23}.
We note that \cite{HoskinsWest_79} uses the SG model driven by an unstable Eady profile to study baroclinic wave development in the atmosphere, simulating its nonlinear evolution before the model's singularity. However, \cite{RagoneBadin_16} point out that \cite{HoskinsWest_79} do not include the lower-order nonlinear term in the PV inversion in SG, which they show to have nontrivial effects.
Our simulation differs from that of \cite{HoskinsWest_79} not only in the choice of the model but also in that we focus on the equilibrium turbulent statistics of the ocean submesoscale by extending the simulation to a large time, achieving a turbulent equilibrium.
The following results show that QG\pl~is able to capture features and statistics commonly associated with the ocean submesoscale.  

The Eady mean state is captured by the mean streamfunction
\begin{align}
    \Phi^M = -\Lambda yz.
\end{align}
It represents the mean westerly sheared velocity ${U}^M = \Lambda z$ and mean buoyancy $ B^M = -f\Lambda y$. We use the mean state to define our Rossby number. The buoyancy advection at the top and bottom boundaries becomes
\begin{subequations}
\begin{align}
    \DD_t b^\text{t}-\Lambda v^\text{t} &= -\mathcal{D}(b^\text{t}),\label{eq:Eady_tbouy_adv}\\
    \DD_t b^\text{b}-f\Lambda H b^\text{b}_x-\Lambda v^\text{b} &= -\mathcal{D}(b^\text{b}),
\end{align}
\end{subequations}
where we have added large and small-scale damping represented by $\mathcal{D}$. 

The inversion relations with the Eady mean state are first derived in \cite{MurakiHakim_01} {(our definition of $F$ and $G$ have a $\Bu$ difference from theirs)}. They are simple modifications of the inversion relation in Section \ref{sec:QGpl_deriv} by noting that with the Eady mean state, the zeroth-order streamfunction becomes $\Phi^0+ \Phi^M$. The QGPV inversion is unchanged because of linearity so in the interior,
\begin{equation}
    \mathcal{L}(\Phi^0) = 0.\label{eq:PV0_inv_Eady}
\end{equation}

All the first-order inversions have modified right-hand sides because of the quadratic nonlinearity. For the first-order PV inversion, we have
{
\begin{align}
    \mathcal{L}(\Phi^1)
    = C_q-\Ub f \left[-\Ub \frac{f^2}{N^2}\Phi^0_{zz}\Phi^0_{zz}-|\nabla\Phi^0_z|^2+2\Lambda\Phi^0_{yz}-\Lambda^2\right]
\end{align}
}
where we have used \eqref{eq:PV0_inv_Eady}. The right-hand side of \eqref{eq:PV0_inv_Eady} must be zero for this simplification. {Since the Eady mean state --- while having zero QGPV --- has non-zero Ertel PV due to the quadratic term ${U}^M_z{B}^M_y$, there is a term of $\Lambda^2$ on the right-hand side}. For the ageostrophic vertical streamfunctions, we have
\begin{align}
    \mathcal{L}\begin{pmatrix}
    F^1\\G^1
    \end{pmatrix} = \Ub 2f\begin{pmatrix}
    J(\Phi_z^0,\Phi_x^0)+\Lambda \Phi^0_{xx}\\
    J(\Phi_z^0,\Phi_y^0)+\Lambda \Phi^0_{xy}
    \end{pmatrix}.\label{eq:Eady_FGinv}
\end{align}

\subsection{Simulation set-up of the QG\pl~model forced by the Eady instability}

{We derive the nondimensional parameters used in our Eady simulation from physical parameters that may very roughly be thought to represent a deep winter mixed layer{ at high latitudes}, with }
{\begin{align*}\label{physscales}
    H &= 250 \text{ m}, \\%\quad 
    \Lambda &= 2.56\times 10^{-4} \text{ s}^{-1},  \\%\quad 
    N &= 8\times 10^{-3} \text{ s}^{-1}\\ %\text{ s}^{-1},\qdt{and}
    f &= 10^{-4} \text{ s}^{-1}.
\end{align*}}
    {Choosing} $\text{Bu} = 1$ sets the length used in the model's nondimensionalization equal to the deformation scale {\citep[cf.][]{dong2020scale}},
\begin{align*}
    {
    L = \frac{NH}{f} = 20\text{ km}.
    }
\end{align*}
{The velocity scale} {$U=\Lambda H = 0.064\text{ m/s}$}, {sets} the Rossby number to $\varepsilon = 0.032$, which characterizes the large-scale Eady mean state.  We will see below that, despite this small value, the emergent local Rossby number $\Ro\zeta/f$ can easily exceed 1, as is found in submesoscale flows. 

We use a horizontally square domain of size $6\pi\times L$, that is {$L_\text{domain}=120\pi\text{ km}$}, so in nondimensional terms, our domain is horizontally periodic and vertically bounded with size $(Lx,Ly,Lz) = (6\pi,6\pi,1)$.  The choice to include only a small scale separation between the domain scale and the deformation scale reflects our intentional focusing of model resolution on resolving submesoscale dynamics in the forward cascade.  

We note a numerical limitation on some parameter choices.  
The total background buoyancy is $N^2z-\Ro f\Lambda y$ and we have found curiously that when either $\varepsilon$ or $L_\text{domain}/L$ is large enough for the background isopycnals to connect the top and bottom surfaces, a large-scale flow instability develops leading to {exceedingly small CFL timestep size requirement, effectively halting simulation progress}. We currently do not have an explanation for this numerical blow-up phenomenon. {But we would like to note that the horizontally doubly periodic set-up is obviously physically unrealizable. {We conjecture that} a channel set-up, more commonly done for primitive equations simulations of the submesoscale, might alleviate this instability. This required imposing horizontal boundary conditions to QG\pl, which is possible but outside the scope of the study.}

The energy injected by the steady baroclincally unstable mean state must be dissipated in order to achieve a turbulent equilibrium. We use the dissipation
\begin{align}
    \mathcal{D}(b) = \nu_0 \overline{b}-\nu_{-2} \nabla^{-2} b+\nu_{4}\nabla^4 b
\end{align}
where all operators are interpreted in the Fourier sense. The inverse Laplacian ($\nabla^{-2}$) term arrests the inverse cascade at large scales. This operator does not act on the constant mode and allows the horizontal average of the buoyancy to drift away from zero (via the mechanism described by \eqref{bbdivg}), which is an important feature of the QG\pl~model. To control this drift, we add a damping term for the constant mode. The value of $\nu_0$ only affects $\overline{b^\text{t,b}}$ and, per our previous comment in Section~\ref{sec:QGpl_deriv}, is inconsequential to the evolution of the model. However, it models important surface buoyancy forcing that restores PV close to the idealized zero PV assumption of the Eady model. We choose $\nu_0$ so that the final turbulent steady-state values of the surface buoyancy $\overline{b^{\text{t},\text{b}}}$ are roughly consistent with \eqref{PVintegral} and $C_q=0$. Finally, we use fourth-order ($\nabla^4$) hyper-dissipation to arrest the frontogenesis at small scales. We choose the {nondimensional} dissipation constants 
\begin{align*}
    \nu_{-2} = 0.35, \quad \nu_0 = 0.34, \qdt{and} \nu_{4} = 6.12\times 10^{-7}. 
\end{align*}

To simulate the three-dimensional QG\pl~system driven by the Eady mean state, we advect buoyancy at the upper and lower surfaces. {Assuming vanishing PV $q=0$, we invert for the four potentials $\Phi^0$, $\Phi^1$, $F^1$, and $G^1$ by solving three-dimensional Poisson problems}. For comparison, we also simulate the the standard QG equations with the same parameters, simply by inverting only for the zeroth-order potential $\Phi^0$. Everything is done pseudospectrally using Dedalus \citep{BurnsEtAl_20}. We use the Fourier basis in the horizontal and Chebyshev in the vertical with the number of modes $(Nx,Ny,Nz) = (512,512,24)$. We check the vertical solution is fine enough for the inversion by running a simulation with 16 and 32 modes and checking that the turbulent statistics are unchanged. Time-stepping is done using the 3rd-order 4-stage diagonally-implicit+explicit Runge-Kutta (DIRK+ERK) scheme \citep{AscherEtAl_97}. The timestep size is determined to have a CFL number of 0.5.

In Appendix~B we show how to simulate the Eady system in a more computationally efficient way, using analytical solutions for the potentials in the interior (similar to the procedure described in Appendix A of HSM02). We choose not to simulate the Eady system this way because we would like our code to generalize to the non-zero PV situation.

We initialize the buoyancy fields with zero-mean small-magnitude random noise and let the instability develop.  The system reaches a statistically steady {state} in about 100 nondimensional time units. We plot snapshots at time $t=190$ and poll the turbulence statistics from the data in the range $t=150-200$.

\subsection{Simulation results of the Eady-forced QG\pl~model}

Eddy kinetic energy (EKE) \eqref{eq:EKE}, available potential energy (APE) \eqref{eq:APE} and their sum the total energy (E) \eqref{eq:total_energy} are common measures of turbulence.  
An alternative measure of energy in our QG-like regime is the energy at the QG-level. Taking only the zeroth-order part of \eqref{eq:TE_potform} gives the familiar QG definition of energy
\begin{align}
    \text{E}^0 &= \left\langle |\nabla \Phi^0|^2+\Ub\left|\frac{f\Phi^0_z}{N}\right|^2 \right\rangle. 
\end{align}

\begin{figure}
    \centering
    \includegraphics{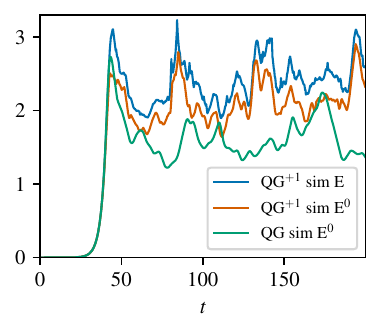}
    \caption{The time evolution of the nondimensional total energies E of the QG\pl~simulation (blue), the QG\pl~simulation at the QG level (E$^0$, red), and of the QG simulation (green). }
    \label{fig:EadyQGpl_fullE}
\end{figure}

\begin{figure}
    \centering
    \includegraphics{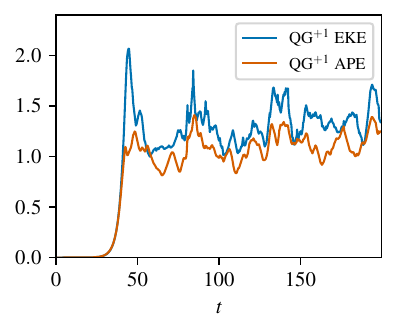}
    \caption{The time evolution of the nondimensional EKE (blue) and APE (red) of the QG\pl~simulation. }
    \label{fig:EadyQGpl_KEPE}
\end{figure}

\begin{figure}
    \centering
    \includegraphics{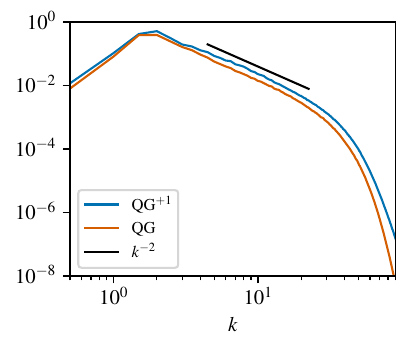}
    \caption{The top buoyancy spectra of the QG and QG\pl~simulation, compared to $k^{-2}$.}
    \label{fig:EadyQGpl_bspec}
\end{figure}

Figure~\ref{fig:EadyQGpl_fullE} shows the total energy (E) and the total energy at the QG level (E$^0$) of the QG\pl~simulation, compared to the total energy of the QG simulation. All of them reach a turbulent steady state after an initial transient. For the QG\pl~simulation, both definitions of total energies are slightly larger than the total energy of the QG simulation, suggesting that the QG\pl~simulation is more vigorous due to ageostrophic frontogenesis.
Figure \ref{fig:EadyQGpl_KEPE} shows the partition of {total energy (E)} between EKE and APE. For our simulation, the variance peaks near the deformation radius. This is evident from the buoyancy spectrum at the top surface in Figure \ref{fig:EadyQGpl_bspec} and for both cases, the submesoscale inertial range has a $-2$ slope, which is similar to high-resolution simulation results \citep{RoulletKlein_10} and matches the prediction of a front-dominated regime \citep{Boyd_92}. The apparent lack of inverse cascade is due to our choice of hypo-dissipation. 

\begin{figure*}
    \centering
    \includegraphics{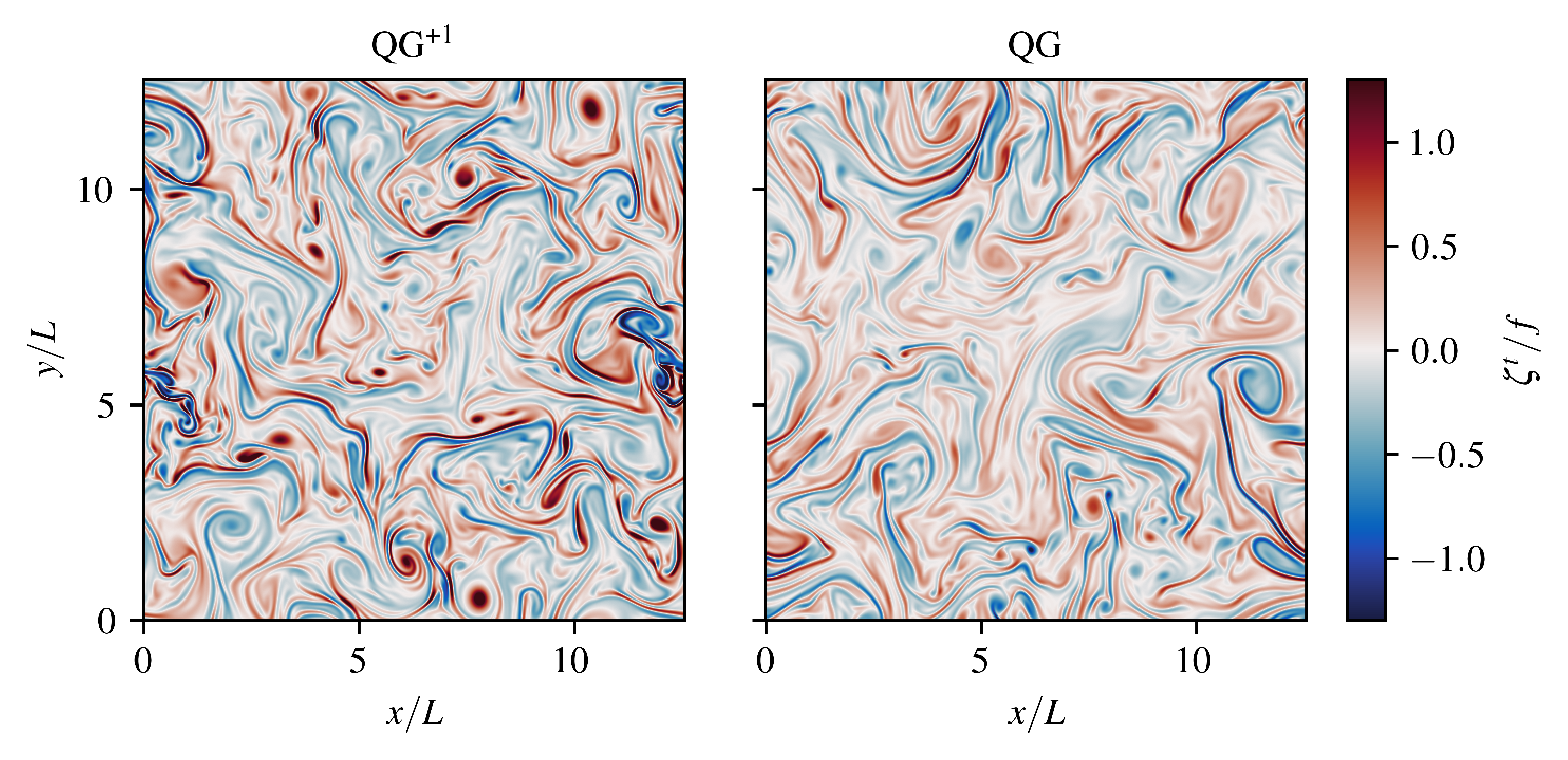}
    \caption{The top nondimensionalized surface vorticity ($\Ro\zeta^\text{t}/f$) at time $t=190$ for the two simulations. The QG\pl~simulation is on the left while the QG simulation is on the right.}
    \label{fig:EadyQGpl_zeta_snap_Ro0d06_t175d0}
\end{figure*}

\begin{figure*}
    \centering
    \includegraphics{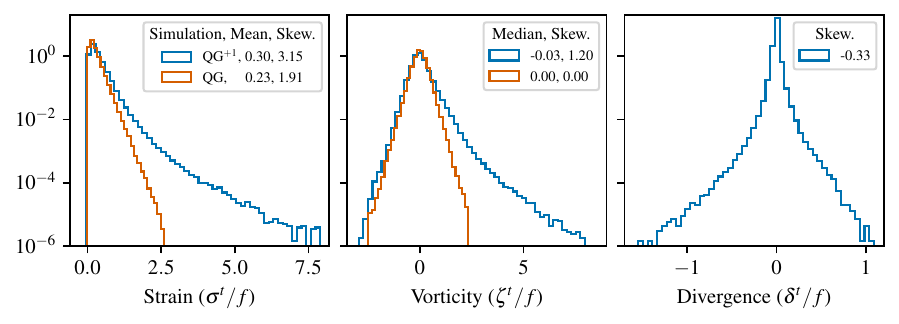}
    \caption{The histograms of turbulent statistics (strain, vorticity, and divergence) for QG (orange) and QG\pl~(blue) simulations polled over $t=150-200$.}
    \label{fig:EadyQGpl_zetaPDFs_Ro0d00}
\end{figure*}

QG\pl~captures turbulent statistics representative of frontogenesis. Ageostrophic effects produce stronger cyclonic surface vorticity, in contrast to QG, which is symmetric in cyclonic and anti-cyclonic features. Figure~\ref{fig:EadyQGpl_zeta_snap_Ro0d06_t175d0} shows the top surface vorticity ($\Ro\zeta^\text{t}/f$) of our simulation. It has stronger and thinner cyclonic vorticity fronts. It also has {small} cyclonic vorticity patches with {fewer and larger} anti-cyclonic ones. Vorticity statistics summarized in the histograms in Figure~\ref{fig:EadyQGpl_zetaPDFs_Ro0d00} confirm that the vorticity is strongly skewed to cyclonic with positive skewness of 1.2, and its median is slightly anti-cyclonic. It is also worth reiterating that the flow-Rossby number ($\Ro\zeta^\text{t}/f$) is indeed $O(1)$, even though the large-scale Rossby number of the mean-field driving the baroclinic instability is small. QG\pl~also captures the stronger strain values ($\Ro\sigma^\text{t}/f$) where
\begin{align}
    \sigma^\text{t} = \left[(u^\text{t}_x-v^\text{t}_y)^2+(v^\text{t}_x+u^\text{t}_y)^2\right]^{1/2}
\end{align}
due to strong fronts, compared to QG (Figure~\ref{fig:EadyQGpl_zetaPDFs_Ro0d00}). 

\begin{figure*}
    \centering
    \includegraphics{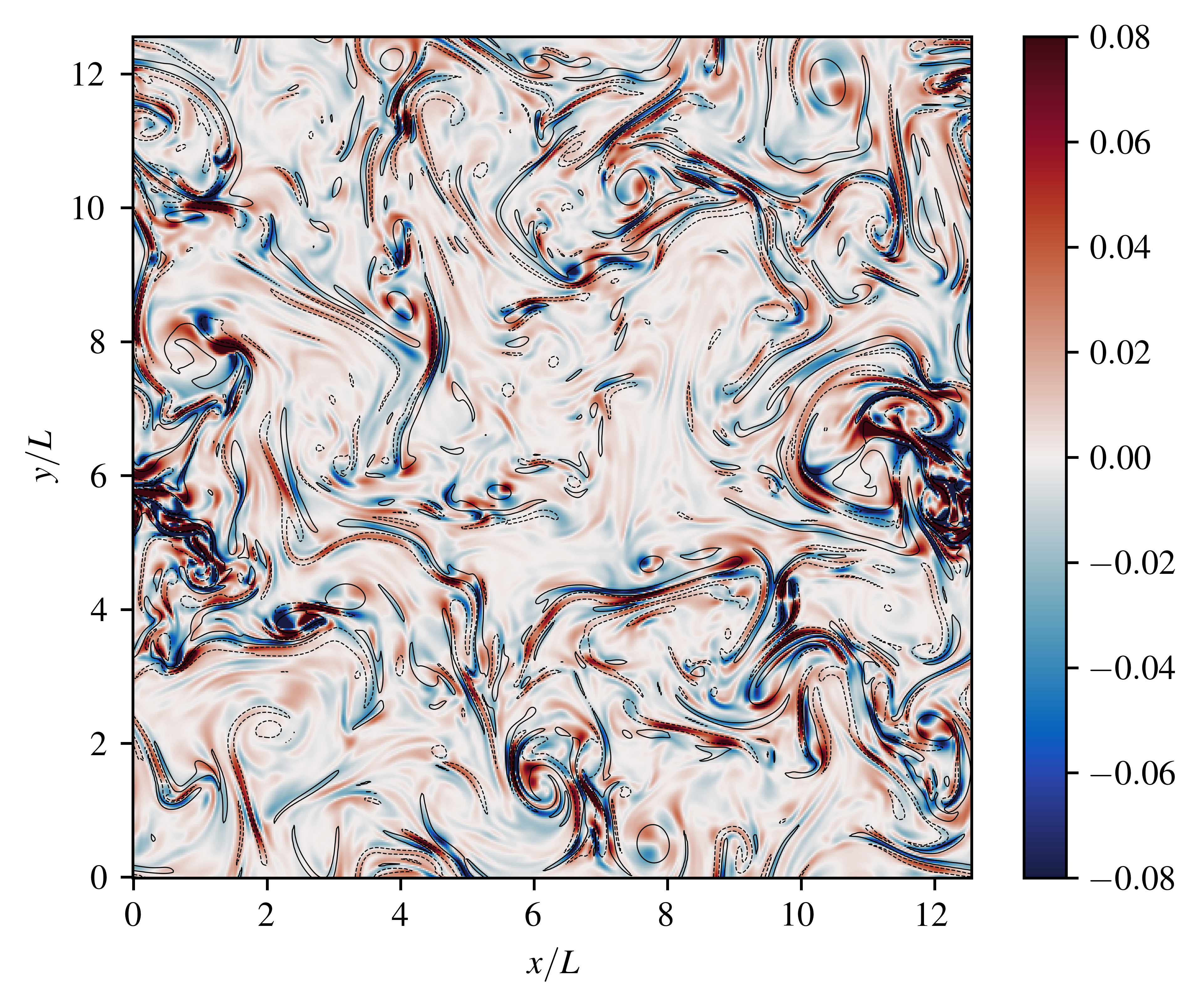}
    \caption{Snapshot of top vorticity {in contour lines} ($\Ro\zeta^\text{t}/f$) and divergence {in color} ($\Ro\delta^\text{t}/f$) of the QG\pl~simulation at time $t=190$. The contours show vorticity levels of $-0.4$ (dashed) and $0.4$ (solid).}
    \label{fig:EadyQGpl_zetadiv_snap_Ro0d06_t175d0}
\end{figure*}

\begin{figure}
    \centering
    \includegraphics{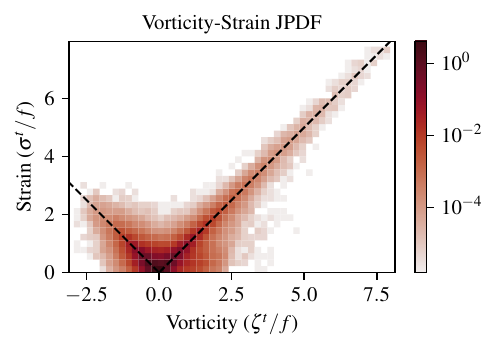}
    \caption{The joint probability density function (JPDF) of top vorticity and top strain for the QG\pl~simulation polled over $t=150-200$.}
    \label{fig:EadyQGpl_jointPDFs_Ro0d03}
\end{figure}

\begin{figure*}
    \centering
    \includegraphics{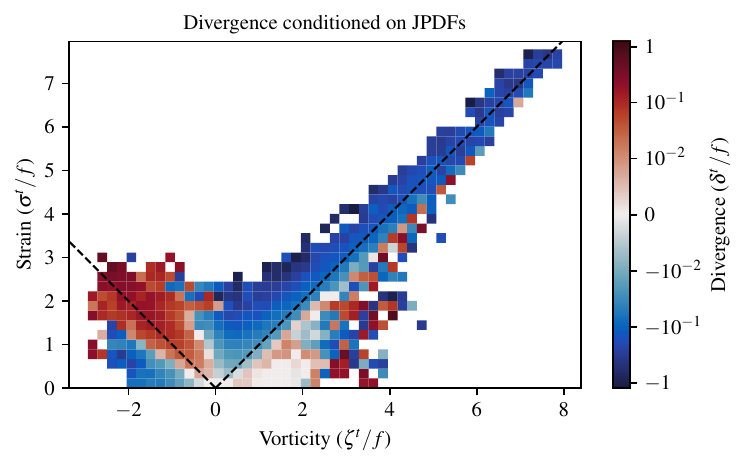}
    \caption{The same joint probability density function (JPDF) as Figure \ref{fig:EadyQGpl_jointPDFs_Ro0d03} except the color is the mean of the top divergence conditioned on the strain and vorticity at that pixel of the histogram.}
    \label{fig:EadyQGpl_condJPDF_Ro0d03}
\end{figure*}

This asymmetry in vorticity can be explained by the non-zero surface divergence captured by QG\pl, correlated with vorticity. In Figure~\ref{fig:EadyQGpl_zetadiv_snap_Ro0d06_t175d0}, top surface divergence ($\Ro\delta^\text{t}/f$) where
\begin{align}
    \delta^\text{t} = u_x^\text{t}+v_y^\text{t}
\end{align}
is in color and top surface vorticity is superimposed as contours. The cyclonic vorticity fronts are correlated with convergence while anti-cyclonic vorticity fronts are correlated with divergence. This matches the expectation of frontogenesis. This is also verified by turbulent statistics in Figure \ref{fig:EadyQGpl_zetaPDFs_Ro0d00}, which confirms $O(1)$ convergence values. Indeed, the joint PDF of strain ($\Ro\sigma^\text{t}/f$) 
and vorticity ($\Ro\zeta^\text{t}/f$) shown in Figure \ref{fig:EadyQGpl_jointPDFs_Ro0d03} display a strong tail in the $x=y$ line. Taking the conditional mean of top divergence in this long tail, we see that it correlates with negative divergence in Figure \ref{fig:EadyQGpl_condJPDF_Ro0d03}. It has been shown from observations and high-resolution primitive equations simulations that this long tail is a signature feature of front-dominated submesoscale oceans that are shrinking because of frontogenesis \citep{ShcherbinaEtAl_13, BalwadaEtAl_21}. 

\begin{figure}
    \centering
    \includegraphics{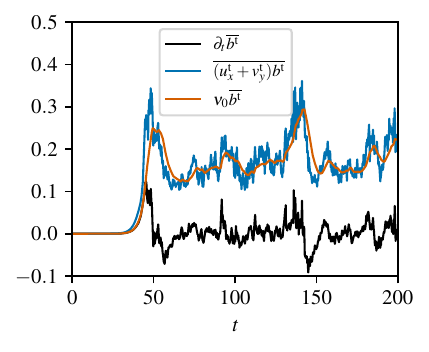}\\
    \includegraphics{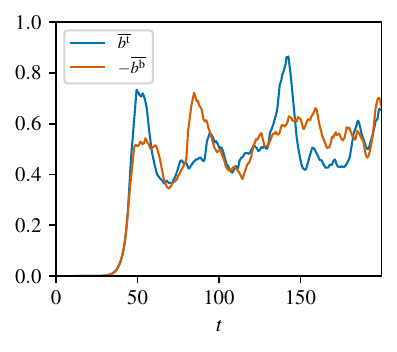}
    \caption{Top: The nondimensional values of the tendency terms of $\overline{b^\text{t}}$ in \eqref{eq:meanb_tend} over the simulation. Bottom: The time evolution of top and (negative) bottom mean buoyancy. }
    \label{fig:EadyQGpl_bmean_Ro0d06}
\end{figure}

QG\pl~also captures the re-stratifying effect of ageostrophic frontogenesis by modeling the correlation between divergence and buoyancy anomaly. The evolution of the mean surface buoyancy \eqref{bbdivg} in the Eady model is
\begin{equation}\label{eq:meanb_tend}
    \pe_t \overline{b^\text{t}} =   \overline{(u^\text{t}_x+v^\text{t}_y)b^\text{t}}+\Lambda \overline{v^\text{t}} -\nu_0 \overline{b^\text{t}}.
\end{equation}
We know $\overline{v^\text{t}} = 0$ and Figure \ref{fig:EadyQGpl_bmean_Ro0d06} plots the other two tendency terms and 
$\pe_t\overline{b^{t}}$ on the top and the time evolution of $\overline{b^\text{t}}$ and $\overline{b^\text{b}}$ on the bottom. The gain in top buoyancy $\overline{b^\text{t}}$ and decrease in bottom buoyancy $\overline{b^\text{b}}$ is driven by the correlation between the divergence of the velocity field and the buoyancy fields \citep{HakimEtAl_02}. Ageostrophic restratification is absent in the QG model since explicit divergence is zero {and the divergence of the ageostrophic
velocity field does not feedback in the evolution of buoyancy}.

\section{QG\pl~and semigeostrophic simulations of {two-dimensional strain-induced} frontogenesis}\label{sec:2D_front}

\begin{figure*}
    \centering
    \includegraphics[scale=0.65]{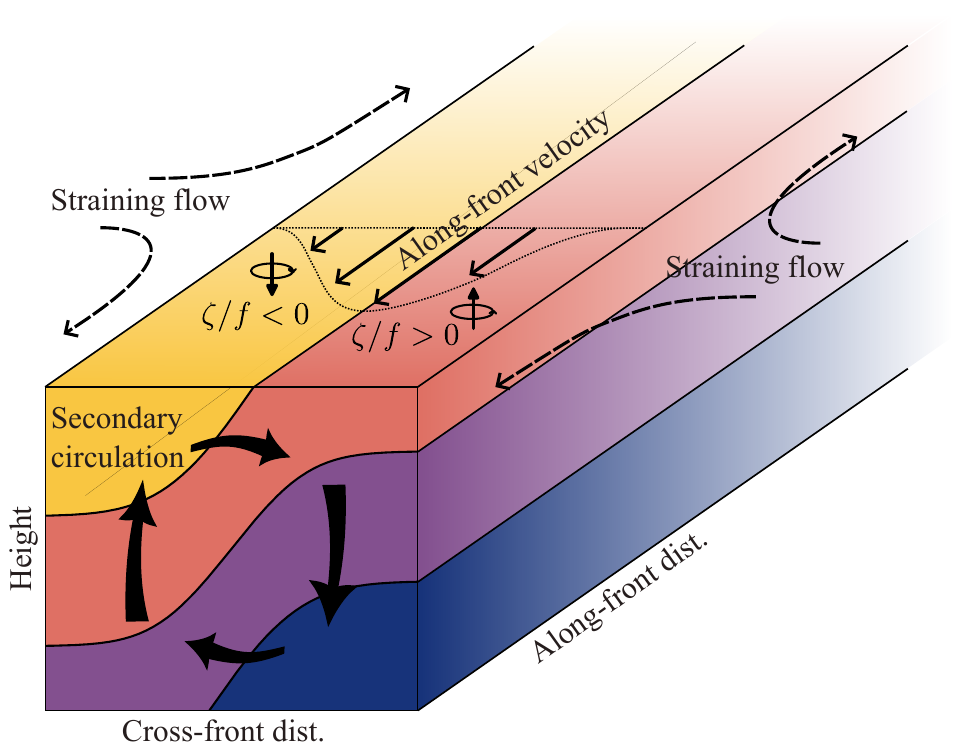}
    \caption{Schematic representation of the strain-induced frontogenesis model of \cite{HoskinsBretherton_72}.}
    \label{fig:front_scheme}
\end{figure*}

To elucidate the QG\pl~model's ability to capture front-dominated ocean submesoscale turbulence statistics, in this section, we zoom in on the evolution of an individual front under the classic \cite{HoskinsBretherton_72} strain-induced frontogenesis set-up, summarized schematically in Figure~\ref{fig:front_scheme}. We show that QG\pl~captures the super-exponential growth of surface buoyancy gradient under the advection of the ageostrophic secondary vertical circulation, with an eventual provable finite time singularity. This recovers the result of the semigeostrophic model. At the same time, QG\pl~is simpler mathematically. The inversion involves simple Poisson problems.

\subsection{The primitive equations description of frontogenesis}

The prototypical strain-induced front is straight, infinitely long, and aligned parallel to the $x$-axis. 
That is, we ignore all variability in the $x$ direction for the response fields. 
The problem is most naturally posed in a vertical bounded domain $z\in[-H,0]$ and infinite in $y$ with decay boundary condition.
It is forced by a horizontal incompressible, irrotational strain field
\begin{align}
    {U}^M=\alpha x, \qquad {V}^M = -\alpha y.
\end{align}
{which} can be captured by a horizontal streamfunction
\begin{align}
    {\Phi}^M = -\alpha xy.
\end{align}
Using the standard mean pressure field $P^M = -\alpha^2(x^2+y^2)/2+f\alpha xy$ to compensate for the straining field, we have the {primitive equations}
\begin{align}
\begin{split}
    \Ro\left(\DD_t u+\alpha u\right)-fv &= 0,\\
    \Ro\left(\DD_t v-\alpha v\right)+fu &= -p_{y},\\
    b &= p_{z},\\
    \Ro\DD_t b+ \Bu wN^2 &= 0, \\
    v_{y}+w_{z} &= 0
\end{split}
\end{align}
with
\begin{align}
    \DD_t = \pe_t+(v+{V}^M)\pe_y+w\pe_z.
\end{align}
Because QG\pl~is based on the QG scaling, we still assume a mean vertical stratification of $N^2$. We use the straining field to define the Rossby number:
\begin{align}
    \varepsilon = \frac{\alpha}{f}.
\end{align}
The horizontal boundary conditions for $u$ and $v$ are decaying at infinity. For $b$, its boundary values are the same as the ones of the initial condition, to maintain the front.
This system has PV conservation
\begin{align}
    \DD_t q = 0
\end{align}
where $q$ is \eqref{PV} without the terms with $x$ derivatives. The irrotational straining field has no effect on the PV. At the boundary, we have horizontal advection ($w=0$) of the surface buoyancy:
\begin{align}
    \DD_t^\text{t} b^\text{t} = 0; \qquad \DD_t^\text{b} b^\text{b} = 0.
\end{align}

\subsection{Frontogenesis modeled by the QG\pl~model}

The QG\pl~inversion follows from the general case after we replace the QG streamfunction with the sum $\Phi^0+{\Phi}^M$. The zeroth order sees no changes from \eqref{eq:PV0_inv}
except we redefine the $\mathcal{L}$ by ignoring the $x$-derivative terms. For the horizontal boundary at infinity, we use the fact that $u\to 0$ as $y\to\pm \infty$ to give
\begin{align}
    \left.\Phi^{0}_y\right|_{y=\pm\infty} = 0.
\end{align}
According \eqref{eq:var_expan}, to obtain $v$ and $w$ for the advection, we only need one ageostrophic secondary streamfunction $G$. Indeed, we have
\begin{align}
    v = -\Ro G^1_z, \quad
    w = \Ro G^1_y.
\end{align}
Now for the inversion of $G$, we have
\begin{align}
    \mathcal{L}(G^1) = 2Q_2
\end{align}
where 
\begin{align}
    Q_2 &= \Ub fJ(\Phi^0_z+{\Phi}^M_z,\Phi_y+{\Phi}^M_y) = \Ub f\alpha\Phi^0_{yz}.
\end{align}
This is the $\omega$-equation in \cite{HoskinsEtAl_78a}.  Note that without the straining field ($\alpha=0$), there would be zero secondary circulation. This is consistent with the fact that the initial condition is in geostrophic balance and is a solution of the {primitive equations} with no strain.We use $v\to 0$ as $y\to\pm y$ to give Dirichlet horizontal boundary condition at infinity
\begin{align}
    \left.G^{1}\right|_{y=\pm\infty} = 0.
\end{align}

We only need $\Phi^1$ for $u$ and $b$:
\begin{align}
    u &= -\Phi^0_y-\Ro\Phi^1_y\\
    b &= f\Phi^0_z-\Ro f\Phi^1_z.
\end{align}
It is a special feature of straight fronts that $u$ and $b$ are always in thermal wind balance in the QG\pl~model. This is the same as in the semigeotrophic model. To invert for $\Phi^1$ we solve:
\begin{align}
    \mathcal{L}(\Phi^1) &= C-\Ub f [\Phi^0_{yy}\Phi^0_{zz}-(\Phi^0_{yz})^2].
\end{align}
The horizontal boundary conditions are
\begin{align}
    \left.\Phi^{1}_y\right|_{y=\pm\infty} = 0.
\end{align}

\subsection{Frontogenesis modeled by the semigeotrophic model}

We would like to compare QG\pl~to the semigeotrophic model under the straight front set-up {to} show that QG\pl~captures the qualitative essence of the semigeotrophic model. We briefly review the semigeostrophic model here. Under the semigeotrophic assumption, $u$ is always in geostrophic balance with the pressure field. That is, we can ignore the $\DD_t v$ term. Under these simplifications the {primitive equations} become
\begin{align}
\begin{split}\label{eq:semi_hydro}
    \DD_t u^g+\alpha u^g-fv^a &= 0,\\
    u^g &= -\phi_{y},\\
    b &= f\phi_{z},\\
    \DD_t b+\Bu w^a N^2 &= 0,\\
    v^a_{y}+w^a_{z} &= 0,
\end{split}
\end{align}
where $\phi = p/f$ is the geostrophic streamfunction and
\begin{align}
    \DD_t = \pe_t +({V}^M+\Ro v^a)\pe_y+\Ro w^a\pe_z.
\end{align}
Here we use superscript $g$ for velocities in geostrophic balance and $a$ for those which are not. 
% This model conserves the simplified Ertel PV \citep{ThomasEtAl_08a}:
% %%\begin{table*}[H]
% \begin{align}
%     Q &= fN^2+\Ro\left(N^2\phi_{yy}+\Ub f^2\phi_{zz}\right)+\Ro^2\Bu \left(f\phi_{yy}\phi_{zz}-f\phi_{yz}^2\right)\label{eq:semi_ErtelPV}\\
%     &= fN^2+\Ro q.\label{eq:semi_PV}
% \end{align}
%%\end{table*}

The elliptic problem for the vertical ageostrophic streamfunction $G^{\text{SG}}$
\begin{align}
    v^a &= -G^{\text{SG}}_z,\qquad w^a = G^{\text{SG}}_y.
\end{align}
is the Eliassen-Sawyer equation \citep{ThomasEtAl_08a}, which can be derived using similar steps in the derivation of the $\omega$-equation
\begin{align}
    \left(\Ro f\phi_{zz}+\Bu N^2\right) G^{\text{SG}}_{yy}-\Ro 2f\phi_{yz} G^{\text{SG}}_{yz}+f\left(f+\Ro \phi_{yy}\right) G^{\text{SG}}_{zz} = 2\alpha f \phi_{yz}.\label{eq:semi_ES_eq}
\end{align}
This is a variable coefficient elliptic problem, which is harder to solve than the constant Poisson problem used in the QG\pl~inversion. The boundary conditions for $G^{\text{SG}}$ are the same {as the} ones in QG\pl.

With this inversion established, we simulate the semigeotrophic model by using buoyancy as the prognostic variable. The buoyancy conservation equation is
\begin{align}
    \DD_t b+\Bu w^a N^2 &= 0.
\end{align}
Given $b$, we can use hydrostatic balance equation \eqref{eq:semi_hydro} to obtain the pressure field $\phi$. From here we can solve the Eliassen-Sawyer equation \eqref{eq:semi_ES_eq} to obtain $v$ and $w$. It is worth remarking that the semigeostrophic model can also be solved by using PV ($q$) as the single prognostic variable. However, the inversion from PV is even harder, requiring solving the nonlinear Monge–Ampère equation \citep{HoskinsBretherton_72,RagoneBadin_16}.

\subsection{Simulation set-up for the two balanced models of frontogenesis}

\begin{figure}
    \centering
    \includegraphics{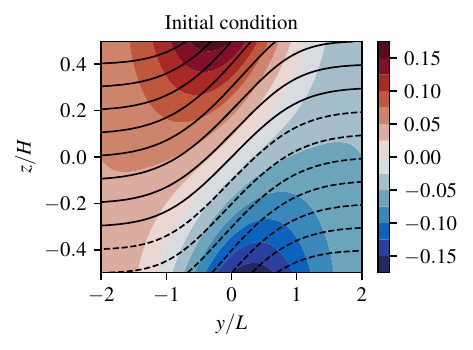}
    \caption{The initial condition of the two-dimensional front simulation. The figure zooms in on the front ($y/L\in[-2,2]$). The color is the along front velocity $\Ro u/\alpha L$ and the contour is buoyancy ${\left.\left(\Ro b+\Bu N^2z\right)\right/N^2H}$. Buoyancy contours have gaps of 0.1 and the dashed line is negative.}
    \label{fig:2DFSemiu_t0}
\end{figure}

We use Dedalus to simulate the strain-induced frontogenesis under the QG\pl~and the semigeostrophic model under large strain ($\varepsilon = 0.3$). We simulate the $q=0$ case. A variable PV case can be simulated with minimal modifications. We start the top and bottom perturbation buoyancy to be
\begin{align}
    b^\text{t}(t=0) = b^\text{b}(t=0) = -\erf(y). 
\end{align}
We can obtain the full two-dimensional buoyancy by solving the QG\pl~inversion. Figure \ref{fig:2DFSemiu_t0} shows the initial buoyancy and the implied along front velocity field $u$. For the QG\pl~model, we advect the PV and surface buoyancy. For semigeostrophy, we advect the buoyancy everywhere. The velocities for the advection come from the inversion procedures described above. 

\begin{figure}
    \centering
    \includegraphics{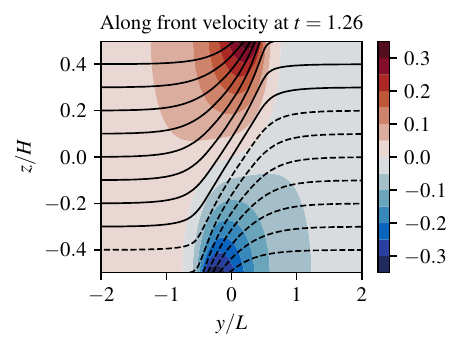}\\
    \includegraphics{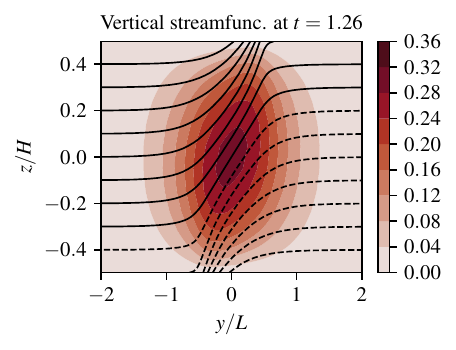}
    \caption{Top: Same as Fig. \ref{fig:2DFSemiu_t0} but for late-stage frontogenesis at time $t=1.26$. Bottom: The color now shows the secondary circulation streamfunction $G/UH$ at the same time.}
    \label{fig:2DF_snap}
\end{figure}

Above we posed the two-dimensional frontogenesis problem on an infinite $y$ domain. However, for numerical convenience, we simulate it on a finite interval and impose the boundary condition in the $y$-walls. The boundary conditions for $b$ are
\begin{align}
    \left.b\right|_{y=-Ly/2} = 1; \qdt{and} \left.b\right|_{y=Ly/2} = -1.
\end{align}
And for PV, we have
\begin{align}
    \left.q\right|_{y=\pm Ly/2} = 0.
\end{align}
It is difficult to simulate dynamics that tend to a singularity. We use the 1st-order 1-stage DIRK+ERK scheme for time-stepping so that the numerical error is purely dissipative \citep{AscherEtAl_97}. We run the simulation with various resolutions and domain sizes to allow for longer simulation before numerical blow-up and check that the finite domain size does not affect our result.

\subsection{Simulation results of frontogenesis}

\begin{figure*}
    \centering
    \includegraphics{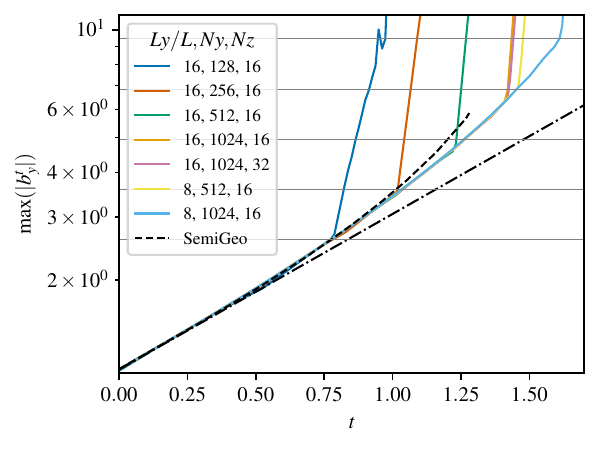}
    \caption{The evolution of maximum nondimensional top buoyancy gradient for the two-dimensional frontogenesis. The solid curves are the results of the QG\pl~model, labeled by the nondimensional horizontal domain size $L_y/L$, and numbers of resolved modes $(N_y,N_z)$. The dashed line is the longest simulation of the semigeostrophic model we are able to obtain.
    The dash-dot line is the exponential growth $1.126e^{t}$.
    The gray horizontal lines are $2.6\times1.38^n$ for $n\in \{0,1,2,3,4\}$. 
    }
    \label{fig:2DFQGpl_growth}
\end{figure*}

Frontogenesis pushes the gradient of surface buoyancy to grow super-exponentially, and to an eventual finite time singularity in the semi-geostrophic model. QG\pl~captures the same phenomenology as shown {in the plot} of the maximum top buoyancy gradient in Figure \ref{fig:2DFQGpl_growth}. The QG\pl~results are the many solid curves labeled by the {sizes} of the domain and the horizontal and vertical {resolutions}. They all become numerically unstable at some point, but before that, they follow a common curve. Additionally, the numerical blow-up times roughly follow a power-law of the resolution {($Ly/Ny$)}, indicating it is due to the lack of horizontal resolution. This curve first grows exponentially following the straining flow{, then} it enters into a super-exponential regime. It is reasonable to extrapolate that the dynamics will reach a singularity at a finite time. Indeed we have theoretical proof {in Appendix C} that it does and the time of blow-up is proportional to the strain. This super-exponential growth is caused by the secondary circulation $G$ (shown in Figure \ref{fig:2DF_snap}) and is a hallmark of ageostrophic frontogenesis. {A} snapshot of the along front velocity $u$, imposed on top of the buoyancy field $b$, at time $t=1.26$, well into the super-exponential phase of frontogenesis is displayed in Figure \ref{fig:2DF_snap}. These plots show a sharp front on the top cold side {which} is the strong convergent region that pushes the growth of the buoyancy gradient. To keep incompressibility, the vertical velocity $w$ is large and grows negative there. We have a front undergoing frontogenesis.

We also simulate the frontogenesis under the semigeostrophic model for comparison. The dashed line in Figure \ref{fig:2DFQGpl_growth} is the growth of the maximum top buoyancy gradient for the longest simulation of the semigeostrophic model we have. We see that the semigeotrophic model grows faster. Otherwise, their growth looks similarly super-exponential. It is harder to simulate the semigeostrophic model for a long time. The numerical blow-up time is no longer monotonic to the horizontal resolution. We suspect this is because the variable coefficient elliptic problem is harder to solve numerically. The snapshots of the evolution look very similar to the snapshots of the QG\pl~model (not shown).

It is indeed impressive that the QG\pl~model can match the result of the semigeotrophic model. But we need to remind the reader that the semigeostrophic model loses accuracy at late frontogenesis. Instead of comparing the details of the two simulations, we want to instead emphasize that QG\pl~is another balanced model that includes the ageostrophic vertical circulation $G$ in the advection. Its inversion is simple, involving only a constant coefficient Poisson problem. This makes it a promising alternative for the study of balanced, and balanced-wave dynamics under frontogenesis, for which the semigeotrophic model is currently the default {balanced model}.

\section{Conclusions and speculations}
\label{sec:conclu}

Through simulating the QG\pl~model under forced-dissipative turbulence steady state forced by the Eady instability, and frontogenesis of a two-dimensional straight front, we have demonstrated that the QG\pl~model could capture the balanced dynamics of the ocean submesoscale beyond the ability of QG. In particular, the QG\pl~model produces strong fronts with $O(f)$ vorticity and convergence with realistic cyclone--anticyclone asymmetry and statistics matching fronts dominated primitive equations simulations. It can also model the super-exponential growth of fronts, matching the result of the classic semigeostrophic model of \cite{HoskinsBretherton_72}, with a provable finite time blow-up. This makes QG\pl~the unique balanced model that captures balanced ageostrophic effect while being straightforward to simulate in three-dimension. It is also asymptotically consistent and can model curved fronts \citep{GentEtAl_94}. {We hope it will prove to be} a valuable model for the study of the submesoscale ocean. 

QG\pl~is a PV balanced model where it only has a single advected variable, PV. The rest of the model consists of inverting for the physical variables via solving elliptic problems. {For physical understanding, QG\pl~simplifies the primitive equations by {removing} the wave motions {\textit{a priori}}. In particular, the three-dimensional simulation of QG\pl~model provides the balanced motion that is responsible for the vertical transport of tracers.} 
{For data assimilation, QG\pl~reduces the number of prognostic variables (from three to one) in the equations, making it a simpler and possibly more stable model to use. For this reason, we believe that QG\pl, in particular, SQG\pl~could be a useful model for reconstructing three-dimensional ocean flows from limited data, like sea surface height snapshots produced by SWOT.}

In summary, the QG\pl~model\ldots 
\begin{itemize}
     \item faithfully captures asymmetries seen in upper ocean submesoscale, including high cyclonic skewness, stronger cold fronts and filaments, and $O(f)$ horizontal divergence \citep{Barkan2019};
     \item {is} formally one higher order in Rossby than SG, allowing it to represent cyclostrophic balance and curvature;
     \item captures finite-time frontal collapse dynamics{, like its cousin SG};
     \item is {governed by the advection of a single active scalar, from which all other fields are generated through elliptic inversions, making the model straightfoward} to solve numerically;
     \item ensures PV conservation at its core, which is arguably more important that energy conservation in realistic forced/dissipative flows.
\end{itemize}
 
We have only simulated two simple idealized set-ups using QG\pl. With these as proof of concept, we believe it is worthwhile to simulate QG\pl~under realistic ocean states and couple the QG\pl~model with other important components of the climate (e.g.: bio-geochemistry, moist atmosphere, and sea ice). Future works should also explore the detailed dependence of model statistics on the Rossby number, its representation of the submesoscale forward energy cascade, and vicious source terms' effects on PV. These simulations will produce a wealth of turbulence phenomena that are beyond the classic QG model simulations.

\clearpage
%%%%%%%%%%%%%%%%%%%%%%%%%%%%%%%%%%%%%%%%%%%%%%%%%%%%%%%%%%%%%%%%%%%%%
% ACKNOWLEDGMENTS
%%%%%%%%%%%%%%%%%%%%%%%%%%%%%%%%%%%%%%%%%%%%%%%%%%%%%%%%%%%%%%%%%%%%%
\acknowledgments
We thank Jiajie Chen and Vlad Vicol for connecting the QG\pl~model of strain-induced frontogenesis to the Córdoba-Córdoba-Fontelos (CCF) model. 
This article is over the JPO word limit. We thank David Marshall and an anonymous reviewer for their commitment and patience in reviewing this paper and providing valuable feedback. This work was supported in part through the NYU IT High Performance Computing resources, services, and staff expertise. We thank Shenglong Wang for his work on configuring Dedalus on the NYU Greene cluster. KSS and RD acknowledge support from NASA contract 80NSSC20K1142.

%%%%%%%%%%%%%%%%%%%%%%%%%%%%%%%%%%%%%%%%%%%%%%%%%%%%%%%%%%%%%%%%%%%%%
% DATA AVAILABILITY STATEMENT
%%%%%%%%%%%%%%%%%%%%%%%%%%%%%%%%%%%%%%%%%%%%%%%%%%%%%%%%%%%%%%%%%%%%%
\datastatement
The code that generate the data and figures in this paper is available on GitHub: \url{https://github.com/Empyreal092/QGp1_EadyTurb_2DFront_Public}. 

%%%%%%%%%%%%%%%%%
%APPENDIXES
%%%%%%%%%%%%%%%%%
%\setcounter{section}{0}
%\renewcommand{\thesection}{\Alph{section}}

\appendix[A]
\appendixtitle{The equivalency of QG\pl~model and the V96 model}
%\section{foo}
%\label{appV96}

The model in Section 4(a) of V96 is equivalent to the QG\pl~model with some specific choice of the constant $C_q$. However, it lacks the detailed treatment of the top and bottom surface buoyancy conditions done in QG\pl. To show equivalency, we just need to show the velocity fields from the two inversions are the same. 

First, we realize that at the QG level, the two models are the same except the V96 model does not specify the integral constraints requirement of the inversion. We assume the two models make the same choices so that $\psi_0$ in V96 is equal to $\Phi^0$ in QG\pl.

For the higher-order corrections, we start with the vertical velocity $w$. V96 solves for $w$ from a three-dimensional Poisson problem that is the usual $\omega$-equation for $w$ (see (73) in V96). This is the same as the QG\pl~model. We can obtain the Poisson equation for $w$ from \eqref{eq:F_inv}, \eqref{eq:G_inv}, and \eqref{eq:var_expan} and they give us the $\omega$-equation. For the horizontal velocities, two-dimensional Helmholtz decomposition tells us that two two-dimensional zero-mean vector fields are equivalent as long as they have the same divergence and vorticity. Divergence is linked to vertical velocity for an incompressible fluid like the Boussinesq system. Therefore the only thing left to show is the equivalency of vorticity. To do this, we show that the QG\pl's~first order correction to the vorticity is equivalent to the V96's. For V96, $\zeta^1 = \nabla^2\psi_1$. The equations for it are (76), (80), and hydrostatic balance. For QG\pl, we have
\begin{align}
\begin{split}
    N^2 \zeta^1 &= N^2(-u^1_y+v^1_x)\\
    &= N^2(\Phi^1_{yy}+F^1_{yz}+\Phi^1_{xx}-G^1_{xz})\\
    &= N^2\nabla \Phi^1 +\Ub f^2\Phi^1_{zz}-\left[\Ub f^2\Phi^1_{zz}+N^2\left(G^1_{xz}-F^1_{yz}\right)\right]\\
    &= N^2\nabla \Phi^1 +\Ub f^2\Phi^1_{zz} -\Ub fb^1.
\end{split}
\end{align}
That is,
\begin{align}
\begin{split}
    N^2 \zeta^1+\Ub fb^1 &= N^2\nabla \Phi^1 +\Ub f^2\Phi^1_{zz}\\
    &= -\Ub[(v^0_x-u^0_y)b^0_z-v^0_zb^0_x+u^0_zb^0_y]
\end{split}
\end{align}
where we ignored $C_q$. This along with hydrostatic balance is equivalent to (76) of V96. Additionally, 
\begin{align}
\begin{split}
    \nabla^2 b^1-f\zeta^1_z &= \nabla^2\left[ f\left(\Phi^1_z+\Bu \frac{N^2}{f^2} G^1_x- \Bu \frac{N^2}{f^2} F^1_y\right) \right]-f(\Phi^1_{yy}+F^1_{yz}+\Phi^1_{xx}-G^1_{xz})_z\\
    &= \Bu \frac{1}{f}\left(N^2\nabla^2+\Ub f^2\pe_{zz}\right)\left( G^1_x-F^1_y\right)\\
    &= 2J(\Phi^0_x,\Phi^0_y)_z.\label{eq:V96_clyO1}
\end{split}
\end{align}
This with hydrostatic balance is equivalent to (80) of V96. 

In summary, the model in Section 4.3 of V96 is equivalent to QG\pl, ignoring the need for integral constraints in the inversion. This is because V96 does not explicitly address the issue of vertical boundary conditions for the inversion. The QG\pl~model, with the vector potential formulation, addresses this oversight and makes it useful to model the surface ocean. Finally, note that V96 inverts for vorticity, which is a derivative quantity and can be less smooth. QG\pl~inverts for the vector potentials, which can be numerically better conditioned. 

\appendix[B]
\label{appSolvingEady}
\appendixtitle{Solving the QG\pl~inversion under the Eady mean state}

We show the solution to the $F^1$ inversion for the horizontally doubly periodic QG\pl~model under Eady mean state. The solution to all other first-order streamfunctions can be derived following the same idea. We have the nondimensional Poisson problem for $F^1$:
\begin{align}
    \nabla^2F^1+\pe_{zz}F^1 = J(\Phi_z^0,\Phi_x^0)+ \Phi^0_{xx}.
\end{align}
with vanishing Dirichlet boundary conditions. Our solution simply extends the solution in Appendix A of HSM02. We have a particular solution
\begin{align}
    F^1_J = \Phi^0_y\Phi^0_z
\end{align}
that takes care of the Jacobian term on the right-hand side. We just need the particular solution $\Phi^1_L$ to the $\Phi^0_{xx}$ term. With these two particular solutions, we can match the boundary condition by solving the Laplace problem with the mismatch as boundary conditions. 

Now to find the particular solution, we first find the solution of
\begin{align}
    \nabla^2\Phi^0+\pe_{zz}\Phi^0 = 0
\end{align}
with boundary condition
\begin{align}
    \left.\Phi^0_z\right|_{z=0} = b^\text{t}; \qquad \left.\Phi^0_z\right|_{z=-1} = b^\text{b}.
\end{align}
Here $b^\text{t}$ and $b^\text{b}$ are the surface buoyancy with zero means. Using the separation of variables and Fourier transform in the horizontal, we have the Fourier coefficient of the solution:
\begin{align}
    \hat \Phi^0(k,\ell,z,t) = \frac{\csch(K)}{K}\left[ \cosh\left(K(z+1)\right)\hat b^\text{t}(k,\ell,t)-\cosh\left(Kz\right)\hat b^\text{b}(k,\ell,t) \right]
\end{align}
where $\hat{\cdot}$ represents the Fourier transform and $K = \sqrt{k^2+\ell^2}$. Using this, we can transform the search for the particular solution into an ordinary differential equation in $z$:
\begin{align}
\begin{split}
    (-K^2+\pe_{zz}) \hat F^1_L(k,\ell,z) &= -2k^2\hat\Phi^0(z)\\
    &= -2k^2 \frac{\csch(K)}{K}\left[ \cosh\left(K(z+1)\right)\hat b^\text{t}(k,\ell,t) -\cosh\left(Kz\right)\hat b^\text{b}(k,\ell,t)\right].
\end{split}
\end{align}
It has the solution:
\begin{align}
    \hat F^1_L(k,\ell,z) = -k^2 \frac{\csch(K)}{K^2}z\left[ \sinh\left(K(z+1)\right)\hat b^\text{t}(k,\ell,t)-\sinh\left(Kz\right)\hat b^\text{b}(k,\ell,t) \right].
\end{align}

\appendix[C]
\label{appBlowUp}

\appendixtitle{Proof of finite time blow-up of a special frontogenesis example of the QG\pl~model}

In this appendix, we study a two-dimensional strain-induced front using the QG\pl~model and show that it blows up in finite time. We modify the case studies in Section \ref{sec:2D_front} by studying the ``surface'' configuration with only one active surface on the top, infinitely deep bottom, and zero PV in the interior. We will show that the nondimensional evolution of the upper surface buoyancy ($b$) under QG\pl~model follows a one-dimensional transport equation with nonlocal velocity:
\begin{align}
    \pe_t b + \Ro\mathcal{H}(b) b_y = 0\label{eq:CCF_simp}
\end{align}
where the ageostrophic circulation velocity depends on $b$ non-locally through a Hilbert transform
\begin{align}
    \mathcal{H}(b)(y) = \frac{1}{\pi}\, \operatorname{p.v.} \int_{-\infty}^{+\infty} \frac{b(y')}{y - y'}\,\de y'
\end{align}
\cite{CordobaBarbaEtAl_05} first studied this model and proved that it blows up in finite time. Therefore it has been referred to as the CCF model. Readers can refer to \cite{SilvestreVicol_15} for more proofs of finite-time blow-up. The behavior of \eqref{eq:CCF_simp} matches our expectation of frontogenesis blow-up. Figure 3 in \cite{CordobaBarbaEtAl_05} shows the evolution of the CCF model from an initial smooth symmetric profile that could be mapped to a cold filament (our $b$ is negative of their solution). The buoyancy develops a sharp cusp and the velocity tends towards a step discontinuity. The blow-up behavior is similar to the one expected from the \cite{HoskinsBretherton_72} semigeostrophic model of strain-induced frontogenesis. One could find a similar near blow-up profile in \cite{McWilliamsEtAl_15}'s cold filament-genesis solution due to the so-called turbulent thermal wind. Additionally, a simple change of variable shows that the time scale of the evolution, and thus the time of blow-up, scales with the inverse of the Rossby number. Remember that the Rossby number is proportional to the strain rate of the background flow $\alpha$. This matches the blow-up time scaling of the semigeostrophic model. 

Now to show that the dynamics indeed follow the CCF model. We use the nondimensional form of QG\pl~and Burger number equals to one for convenience:
\begin{align}
\begin{split}
    &\Phi^0_{yy}+\Phi^0_{zz} = 0,\\
    &\qdt{with} \left.\Phi_z^0\right|_{z=0} = b,\quad \left.\Phi_z^0\right|_{z\to -\infty} = 0\\
    &G^1_{yy}+G^1_{zz} = 2\Phi^0_{yz},\\
    &\qdt{with} \left.G^1\right|_{z=0} = 0,\quad \left.G^1\right|_{z\to -\infty} = 0\\
    &v = -\Ro G^1_z.
\end{split}
\end{align}
Given the Fourier transform of the one-dimensional $b(y)$ to be $\hat b(\ell)$, we seek the Fourier coefficients of $v(y)$. We solve for the Fourier transform of $\Phi^0$:
\begin{align}
    \hat\Phi^0(\ell) = \frac{\hat b}{|\ell|} e^{i\ell y}e^{|\ell| z}.
\end{align}
For $G^1$, we assume the ansatz 
\begin{align}
    \hat G^1(\ell) = e^{i\ell y}\left[ Ae^{|\ell|z}+Bze^{|\ell|z} \right].
\end{align}
The vanishing Dirichlet boundary condition demands $A = 0$. We solve the elliptic problem to get
\begin{align}
    B = {i\hat b}\sgn(\ell).
\end{align}
Finally
\begin{align}
    \hat v(\ell) = -\Ro\hat G^1_{z}(z=0) = -\Ro{i}\sgn(\ell)\hat b(\ell)
\end{align}
which is $\Ro$ times the Fourier representation of the Hilbert transform.
Together with the straining velocity we have the advection of surface buoyancy follows
\begin{align}
    \pe_t  b + \left[\Ro\mathcal{H}(b)-y\right]b_y = 0\label{eq:CCF_wstrain}
\end{align}
Since the straining velocity does not affect whether the model will form similarity in finite time, we can ignore it for the study of blow-up. 

It is conjectured that the CCF model has vanishing viscosity solutions that are $1/2$-Hölder continuous for all time \citep{SilvestreVicol_15}. This implies that the solutions will have a surface buoyancy spectrum of $|k|^{-3}$. This is confirmed by preliminary simulations (not shown). This might be surprising when we expect front-dominated ocean surfaces to have a density spectrum of $-2$ (see Figure \ref{fig:EadyQGpl_bspec}). However, this is not a contradiction. Analysis of the pure CCF model ignores the straining flow component of the velocity, which is mesoscale and can be captured by QG. This straining flow itself can create a flatter $-2$ density spectrum, as is shown in the QG--QG\pl~spectra comparison in Figure \ref{fig:EadyQGpl_bspec}. The addition of next-order corrections has minimal impact on the buoyancy spectra. 

%%%%%%%%%%%%%%%%%
%REFERENCES
%%%%%%%%%%%%%%%%%
\bibliographystyle{ametsocV6}
\bibliography{citation}

\end{document}